# Low-Temperature and High-Energy-Resolution Laser Photoemission Spectroscopy

Takahiro Shimojima[1], Kozo Okazaki[2], and Shik Shin[2]

1 *Quantum-Phase Electronics Center (QPEC) and Department of Applied Physics,*
*The University of Tokyo, Bunkyo, Tokyo 113-8656, Japan*
2 *Institute for Solid State Physics (ISSP), The University of Tokyo, Kashiwa, Chiba 277-8581, Japan*

We present a review on the developments in the photoemission spectrometer with a vacuum ultraviolet laser at Institute for Solid State Physics at the University of Tokyo. The advantages of high energy resolution, high cooling ability, and bulk sensitivity enable applications with a wide range of materials. We introduce some examples of fine electronic structures detected by laser photoemission spectroscopy and discuss the prospects of research on low-transition-temperature superconductors exhibiting unconventional superconductivity.

Table of contents

## 1. Introduction

Photoemission spectroscopy (PES) is a useful technique that is based on the photoelectric effect originally detected by Hertz in 1887[1] and explained by Einstein in terms of the quantum nature of light in 1905.[2] The basic principles of the photoemission process are illustrated in Fig. 1.[3] When light is irradiated on a sample, electrons with a kinetic energy less than the photon energy $h\nu$ are ejected into the vacuum. The binding energy of electrons in a solid is given by the fundamental photoelectric equation $E_B = h\nu - E_{kin} - \phi$, where $E_{kin}$ is the kinetic energy and $\phi$ is the work function of the material. The number of electrons and their kinetic energy reflect the occupied electronic states of the material[3]. During the 1950s and 1960s, major progress in the experimental method, such as the use of an intense X-ray source and a high-resolution photoemission analyzer, was made by K. Siegbahn, allowing the detailed study of solid-state electronic structures[4]. In the 1970s, ultraviolet photoemission spectroscopy (UPS) was established as a separate technique from PES with X-rays (XPS) owing to the development of a discharge lamp using 21.2 and 40.8 eV helium radiation[5]. While XPS provides information about the core-level states at higher binding energies, UPS is used for the investigation of the valence band states. However, the concepts of XPS and UPS were merged after the development of synchrotron radiation, which offers the advantages of a wide range and tunability of the photon energy.[6]

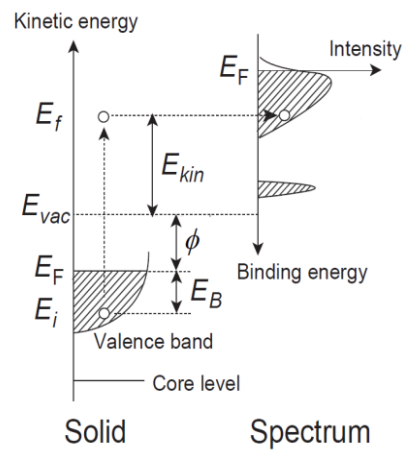

Fig. 1. Schematic view of the photoemission process.

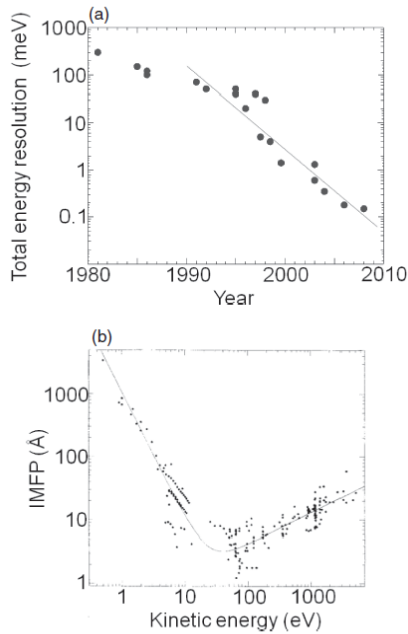

Fig. 2. (a) History of the total energy resolution of PES. (b) Universal curve for the energy-dependent electron inelastic mean free path. Reprinted from Ref. 15.

Improving the energy resolution has been a key issue in boosting the use of PES in research as a sophisticated spectroscopic technique in solid-state physics. Shown in Fig. 2(a) is the history of the highest reported total energy resolution of PES systems. One can find that the energy resolution was rapidly improved after the late 1980s.[7,8] The discovery of high-transition-temperature ($T_c$) copper oxide superconductors (cupra tes) in 1986[9] increased the motivation to improve the performance of PES systems, since the detection of superconducting (SC) gaps requires an energy resolution of 10 meV order. Together with the angle-resolved technique for PES (ARPES), mapping of the SC gap magnitude in a momentum space successfully revealed the $d_{x2-y2}$ pairing symmetry of Cooper pairs.[10] In the last 30 years, the energy resolution has been continuously improved by three orders of magnitude. One of the motivations for achieving high energy resolution has been the detection of the SC gaps of conventional metals with $T_c$ below 10 K. Such observation was required to underpin the understanding of the unconventional superconductivity in cuprates. Furthermore, investigations of the low-$T_c$ superconductors, such as heavy-fermion (HF) systems and organic conductors, are also aimed at revealing the mechanisms of the exotic superconductivity.

To achieve high energy resolution, great effort has been made toward the development of photon sources. The use of a monochromator is an advantage for laboratory sources, *i.e.*, a full width at half maximum (FWHM) of ~1 meV for a helium discharge lamp and bette r than 1 eV for an Al $K\alpha$ or Mg $K\alpha$ X-ray source. The combined improvements in the electron storage ring, undulators, and monochromator enable synchrotron-based PES with energy resolution of less than 1 meV for a low photon energy (~6 eV) and 50 meV for soft X-rays (1–5 keV).[11] In the early 2000s, we started applying a vacuum ultraviolet (VUV) laser to high-energy resolution PES (laser-PES).[12,13] The advantages of the VUV laser include a narrow linewidth, high photon flux, and bulk sensitivity. The bulk sensitivity is shown by the universal curve of the inelastic mean free path (IMFP) in a solid.[14,15] The IMFP at kinetic energies between 20 and 1000 eV is about 5 to 20 Å, and below 20 eV or above 200 eV, this value increases, which is indicative of bulk sensitivity [Fig. 2(b)]. However, there are two difficulties in realizing a laser-PES system: one is the limitation that the photon energy has to be high er than the work function of the specimen (typically 4–5 eV), and the other is the space-charge effect, which broadens the energy of the photoelectrons due to Coulomb repulsion in a dense electron packet.[16] To overcome these issues, a high-repetition-rate (low-peak-intensity) VUV laser of 3.497 eV was employed to obtain the second-harmonic wa ve (6.994 eV) through the nonlinear optical crystal $KBe_2BO_3F_2$ (KBBF).[17,18] The linewidth of the 6.994 eV laser can be reduced to the order of 100 μeV.

The development of a hemispherical electron analyzer has also contributed to the high energy resolution. The mechanical and electronic parameters of the hemispherical electron analyzer and electron lens have been adjusted for use with low-energy photons with a small spot size. As the energy scale of the observation becomes smaller, more advanced cryogenic techniques will be required to remove the thermal broadening effect. For example, the detection of a spectral feature of 1 meV requires a sample temperature ($T$) of less than 10 K. Observations of the SC gap for simple metals (Nb and Pb, $T_c$ of 9.2 and 7.4 K, respectively) in 2000[19] and a low-$T_c$ unconventional superconductor ($KFe_2As_2$, $T_c$ of 3.4 K) in 2012[20] demonstrated that high-energy-resolution PES shows an ongoing growth that will enable us to reveal as-yet-unknown fine electronic structures. Since most physical properties in a bulk and a surface are dominated by the electrons near the Fermi level ($E_F$), the development of a high-energy-resolution PES system has been one of the most important areas of progress in the field of solid-state physics.

In this review, we give detailed descriptions of the first-generation laser-PES system, which has been used since 2005 with energy resolution as fine as 360 μeV and $T$ as low as 2.9 K[12,13], and the second-generation laser-PES system, which started operation in 2012 with performances of 70 μeV and 1.5 K.[20] We also introduce some applications of laser-PES for a wide range of materials, including low-$T_c$ superconductors.

## 2. Development of Laser Photoemission Spectroscopy System

Shown in Figs. 3(a) and 3(b) are a layout sketch and a photograph of the entire first-generation laser-PES system, respectively. The laser-PES system is mainly composed of three parts: a laser system, a cryostat, and an electron analyzer.

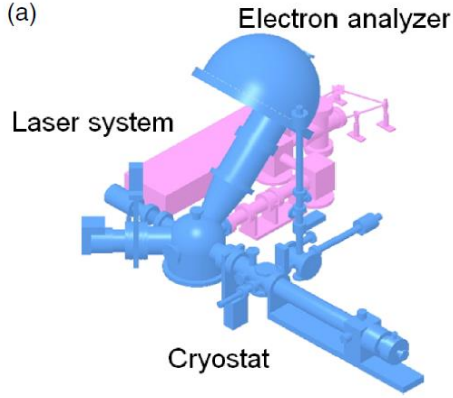

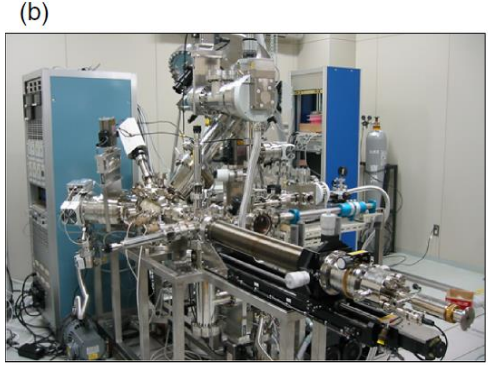

Fig. 3. (Color online) (a) Layout sketch and (b) photograph of the first-generation laser PES system.

These components are attached to an ultrahigh-vacuum measurement chamber, a preparation chamber, and a sample load lock chamber. The base pressure of the measurement chamber is lower than $2.0 \times 10^{-11}$ Torr. As shown in the schematic illustration in Fig. 4, the PES spectrometer and the laser systems are separated by a CaF$_2$ window, through which the 6.994 eV laser light can be transmitted. In this section, we introduce how the high energy resolution and low sample $T$ were achieved in the first-generation laser-PES system.

*2.1 High energy resolution*

The total energy resolution of the PES system $\Delta E_{tot}$ is given by

$$\Delta E_{tot} = \sqrt{(\Delta E_{ana})^2 + (\Delta E_{h\nu})^2 + (\Delta E_{etc.})^2}, \quad (1)$$

where $\Delta E_{ana}$ represents the energy resolution of the electron analyzer, $\Delta E_{h\nu}$ is the FWHM of the photon source, and $\Delta E_{etc.}$ represents external factors such as perturbations of the power supply voltage and grounding. Equation (1) indicates that all of these elements should be comparably reduced in order to obtain a lower $\Delta E_{tot}$. When experimentally estimating actual values of $\Delta E_{tot}$ and sample $T$, a fitting procedure for the Fermi edge cutoff has been implemented by using a Fermi-Dirac (FD) distribution function convoluted with the Gaussian corresponding to the total energy resolution. For the first-generation laser-PES system, a total energy resolution of 360 µeV was achieved at a $T$ as low as 2.9 K, as demonstrated by the PES results for gold as shown in Fig. 5. Here we introduce how the VUV laser system and the electron analyzer contributed to the high energy resolution of the laser-PES system.

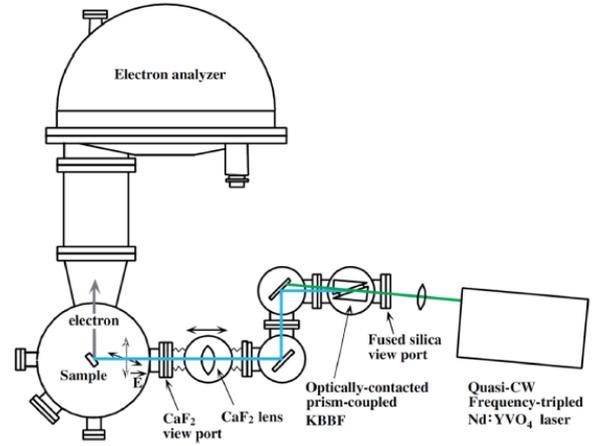

Fig. 4. (Color online) Schematic illustration of the photoemission spectrometer with a VUV laser as a photon source. Reprinted from Ref. 12.

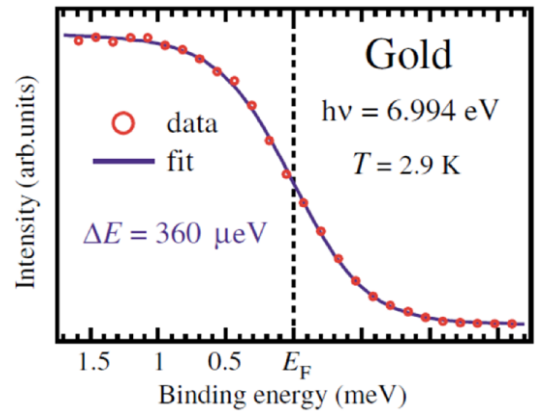

Fig. 5. (Color online) Demonstration of sub-meV resolution of the first-generation laser PES system. The red open circles and blue curve represent the data of gold and the fitting curve, respectively. Reprinted from Ref. 12.

*2.1.1 Vacuum ultraviolet laser system*

The application of a laser to PES has a considerable history, but it has undergone a series of improvements to make it suitable for high-energy-resolution PES.[21,22] A photon source for high-energy-resolution PES requires the conditions of a narrow linewidth, high photon flux, and photon energy higher than the work function of the specimen. While a VUV laser is suitable for these requirements, there is a serious problem called the space-charge effect that broadens the energy of the photoelectrons due to Coulomb repulsion in a dense electron packet.[16] In order to overcome this issue, we selected a frequency-tripled Nd:YVO$_4$ laser (Paladin, Coherent) to generate 355 nm laser light with a high repetition rate of 80 MHz, resulting in a low peak intensity. Figure 6 shows a schematic illustration of the VUV laser system. A 355 nm laser passes through the KBBF crystal at the phase-matching angle of 66.2°, and a second-harmonic wave of 177 nm is then generated. The linewidth of the 177 nm laser is 260 μeV, which is assumed to be a transform-limited Gaussian pulse of 10 ps duration from the Nd:YVO$_4$ laser.[18] The laser output gives a high flux of $2.2 \times 10^{15}$ photons/s for an average output power of 2.5 mW.[18] In order to avoid absorption by the oxygen gas, an optical path for the 177 nm laser is mounted in a vacuum chamber. The laser chamber is connected to the measurement chamber through the CaF$_2$ window to maintain the ultrahigh vacuum in the measurement chamber (~$10^{-11}$ Torr). By using both λ/2 and λ/4 plates, we can choose linear (p or s) or circular (σ+ or σ−) polarizations without changing the optical path. The 177 nm laser is finally focused on the sample. The specifications of the VUV laser and the helium discharge lamp (VG-Scienta VUV5000) are compared in Table I. By employing 6.994 eV photons, photoelectrons are detected from $E_F$ to the binding energy of ~2 eV, and from the surface to the electron escape depth of ~100 Å.[15] In this sense, the VUV laser is a specialized photon source for probing the bulk electronic structure close to $E_F$ with especially high energy resolution.

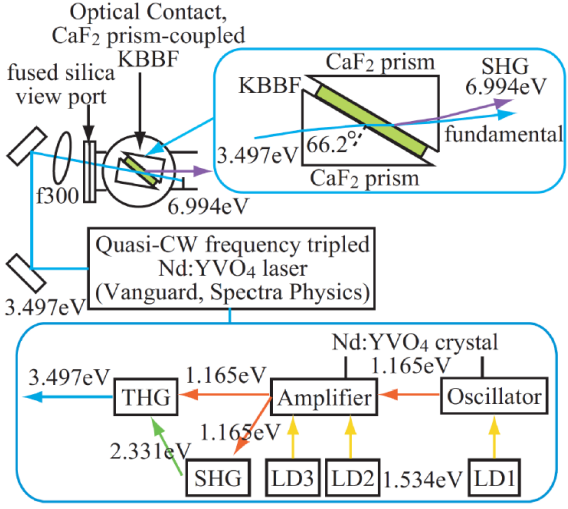

Fig. 6. (Color online) Schematic illustration of the VUV laser system. Reprinted from Ref. 13.

Table I. Specifications of the vacuum ultraviolet laser and the helium discharge lamp

|  | VUV laser | helium discharge lamp |
| --- | --- | --- |
| Spot size | ~200 μm | ~3 mm |
| Photon energy | 6.994 eV | 21.218 eV (He Iα) |
| FWHM | ~260 μeV | ~1.1 meV |
| Photon flux | $2.2 \times 10^{15}$ photons/s | ~$10^{13}$ photons/s |
| Repetition ratio | 80 MHz | Continuous wave |
| Polarization | p, s, σ+, σ− | N/A |

*2.1.2 Electron analyzer*

The VG-Scienta R4000 electron analyzer was developed for the laser-PES system with an energy-resolving power $E_p/\Delta E_{ana}$ of 4000, where $E_p$ is the pass energy of the electron analyzer. The energy resolution of the electron analyzer $\Delta E_{ana}$ is approximated as

$$\Delta E_{ana} = \frac{\omega E_p}{2r_0}, \quad (2)$$

where $\omega$ represents the slit width and $r_0$ is the radius of the hemisphere (Fig. 7). Table II shows a comparison of these parameters between the R4000 and SES2002 electron analyzers in our laboratory. According to Eq. (2), the best performance of R4000 is theoretically estimated to be $\Delta E_{ana}$ of 250 μeV. An electrical lens system with a wide acceptance angle of 38° has been used with an angular resolution of ±0.05° being obtained, corresponding to a momentum resolution of ±0.0008 Å$^{-1}$, using a 6.994 eV laser and a typical work function of 4.3 eV for gold.[13] The lens parameters for low-energy photoelectrons have been optimized for large (2 mm) and small (0.2 mm) spot sizes to give the maximum energy and angular resolution. In order to remove the influence of geomagnetism on photoelectrons with low kinetic energy, the magnetic field at the sample position should be kept below 5 mG. The measurement chamber and hemispherical electron analyzer are therefore protected by double μ-metal shields. According to Eq. (1), the ideal $\Delta E_{tot}$ is estimated to be ~360 μeV, taking $\Delta E_{ana}$ of 250 μeV, $\Delta E_{hv}$ of 260 μeV, and negligible $\Delta E_{etc.}$ into account. This value is comparable to that obtained by laser-PES (Fig. 5).

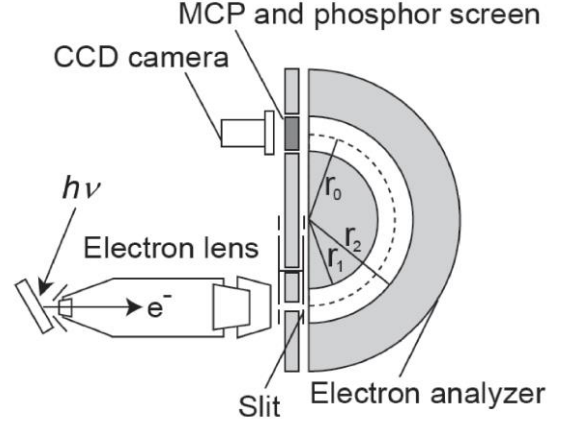

Fig. 7. Schematic illustration of the hemispherical electron analyzer.

Table II. Parameters of the hemispherical electron analyzer.

|  | R4000 | SES2002 |
| --- | --- | --- |
| Pass energy, $E_p$ | ≥ 1 eV | ≥ 1 eV |
| Slit width, $\omega$ | ≥ 100 μm | ≥ 200 μm |
| Radius of analyzer, $r_0$ | 200 mm | 200 mm |
| Energy resolution, $\Delta E_{ana}$ | ≥ 250 μeV | ≥ 500 μeV |

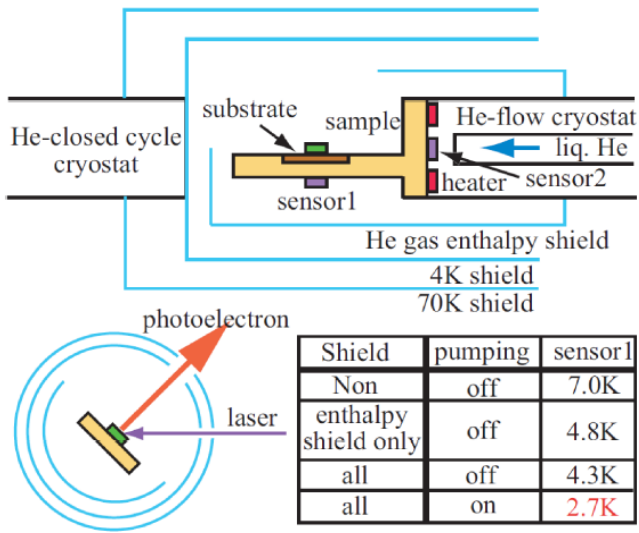

Fig. 8. (Color online) Design of the thermal radiation shields. The inset table shows the lowest $T$ under different cooling conditions. Reprinted from Ref. 13.

*2.2 Cooling system*

In the PES measurements with high energy resolution, the sample $T$ should be kept lower than the energy scale of the total energy resolution. For example, an energy resolution of 1 meV requires a sample $T$ of lower than 10 K. Here we list some issues on which we focused to achieve a low sample $T$.[13] (1) We developed a liquid $^4$He flow-type cryostat with a large cooling power (> 10 W at 4.2 K). (2) We used multiple thermal radiation shields. (3) To ensure high thermal conductivity, the samples were screwed into the hole of the sample holder. (4) The thermal capacity of the sample holder was minimized. (5) Helium gas from the cryostat was pumped out to obtain maximum cooling efficiency. The sample $T$ was monitored with an accuracy of ±0.05 K by using two silicon-diode sensors that were mounted near the sample position and inside the cryostat. The sample $T$ was also controlled by a proportional-integral-derivative algorithm using a heater inside the cryostat with a stability of ±0.01 K.

The thermal radiation shields consist of triple cylinders made of gold-plated copper, *i.e.*, the inner "He gas enthalpy shield," the intermediate "4 K shield," and the outer "70 K shield" as illustrated in Fig. 8. The cylindrical radiation shields have small holes to allow the propagation of the incident light and emitted photoelectrons. To ensure thermal contact, thin gold films were inserted between all mechanical joints in the cryostat. Helium pumping is a useful technique to bring the sample $T$ down to the lambda point (2.2 K) at a reduced liquid helium flow rate of ~0.1 L/h at 5 K in our system. We constructed the helium pumping system using a magnetically coupled rotary pump with a flow rate of 600 L/min. The inset table in Fig. 8 shows the improvement of the lowest $T$ achieved under different cooling conditions. As a result, a minimum $T$ of 2.7 K has been achieved for the first-generation laser-PES system.

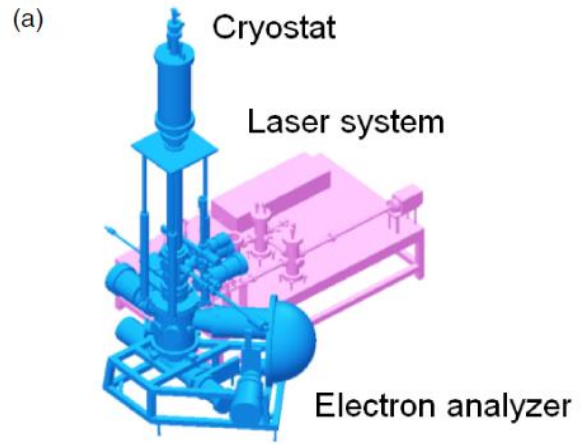
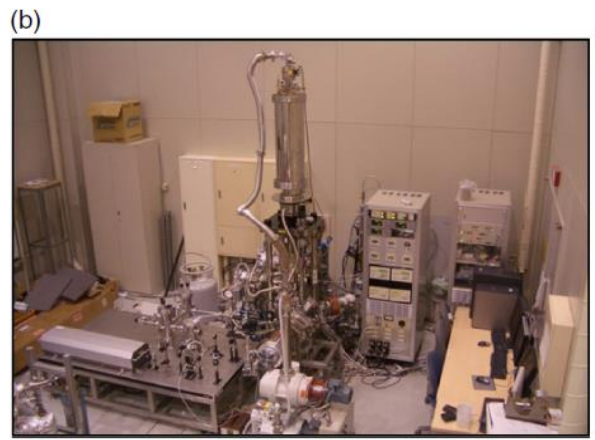

Fig. 9. (Color online) (a) Layout sketch and (b) photograph of the second-generation laser PES system.

*2.3 Second-generation laser photoemission spectroscopy system*

In order to further improve the total energy resolution and the cooling ability, we constructed a second-generation laser-PES system as shown in Figs. 9(a) and 9(b). We developed a vertical-type cryostat with a liquid $^4$He tank surrounded by a liquid nitrogen tank. Liquid $^4$He is transferred to a small room near the sample position and pumped by a rotary pump. In order to reduce the thermal radiation from the measurement chamber and electron analyzer at room temperature, the thermal shields surrounding the samples were further improved. In the VUV laser system, an etalon was introduced to reduce the linewidth of the fundamental laser, which is mounted on a thermostat stage in a chamber filled with dry nitrogen gas. After employing an etalon with a reflectivity of 80%, the linewidth of the second-harmonic wave (6.994 eV) was reduced to 25 μeV. For the hemispherical electron analyzer, we used a VG-Scienta HR8000 and calibrated it with a slit of 50 μm and pass energy of 0.5 eV. The disturbance of the voltage supply and grounding was also reduced to 20 μeV.

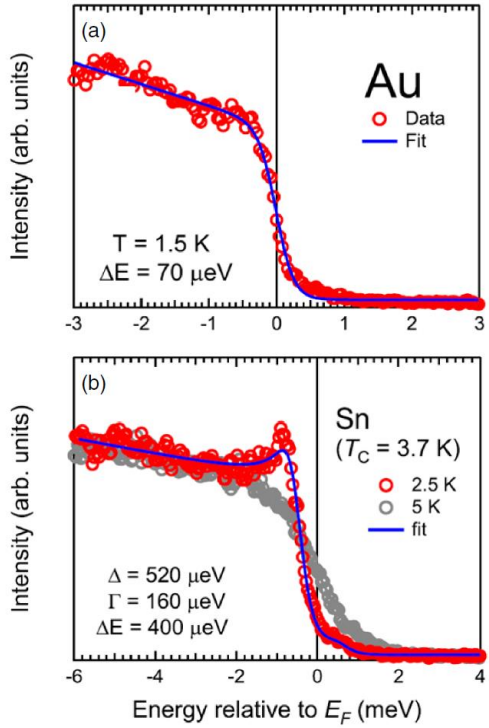

Fig. 10. (Color online) (a) AIPES spectrum of gold obtained by the second-generation laser PES system, with the highest energy resolution of 70 μeV occurring at $T$ = 1.5 K. (b) AIPES spectra of Sn ($T_c$ = 3.7 K) below (red circles) and above $T_c$ (gray circles). The blue curves in (a) and (b) represent the fitting curves based on the FD distribution function and Dynes function,[24] respectively. Reprinted from Ref. 20.

## 2.4 Performance evaluation and target materials

After making these improvements, the highest energy resolution and the lowest $T$ were evaluated from the sharpness of the Fermi edge cutoff of gold. The data in Fig. 10(a) were obtained at the lowest $T$ with the maximum-energy-resolution setup, *i.e.*, a pass energy of 0.5 eV and a slit of 50 μm. By fitting the data with the FD function convoluted by the Gaussian function corresponding to the experimental energy resolution, we obtained a sample $T$ of 1.5 K and a total energy resolution of 70 μeV. The latter value is comparable to $\Delta E_{tot}$ estimated from Eq. (1) using the values of $\Delta E_{ana}$ = 62.5 μeV, $\Delta E_{h\nu}$ = 25 μeV, and $\Delta E_{etc.}$ = 20 μeV.

To demonstrate the high performance of the second-generation laser-PES system, we show the observations of the SC gaps of Sn ($T_c$ = 3.7 K) and Nb ($T_c$ = 9.2 K). In the case of ARPES measurements, the SC gap magnitude can be directly extracted from the fitting procedure using a BCS spectral function of the form

$$A_{BCS}(k,\omega) = \frac{1}{\pi}\left\{\frac{|u_k|^2 \Gamma}{(\omega - E_k)^2 + \Gamma^2} + \frac{|v_k|^2 \Gamma}{(\omega + E_k)^2 + \Gamma^2}\right\}, \quad (3)$$

where $E_k$, $|u_k|^2$, and $|v_k|^2$ are the quasiparticle (QP) energy and the coherence factors of Bogoliubov quasiparticles (BQPs), respectively.[23] By using the normal-state dispersion $\varepsilon_k$ with respect to the chemical potential $\mu$ and SC gap $\Delta(k)$, $E_k$ is expressed as

$$E_k = \sqrt{(\varepsilon_k - \mu)^2 + |\Delta(k)|^2}. \quad (4)$$

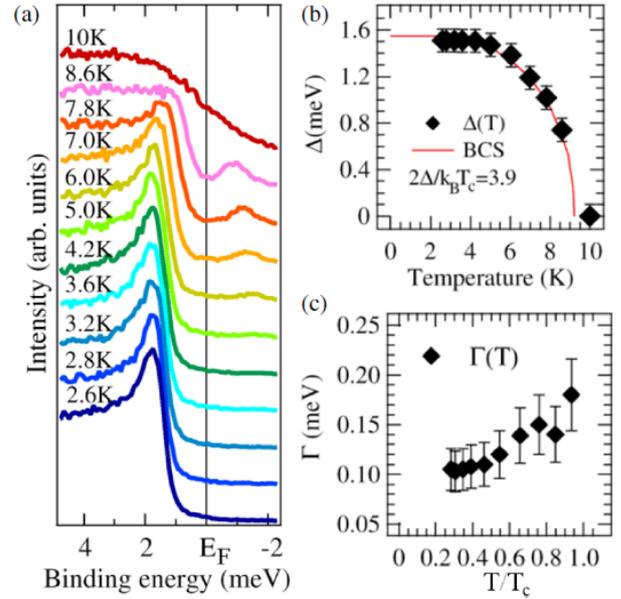

Fig. 11. (Color online) (a) $T$ dependence of the SC gap spectra of Nb ($T_c$ = 9.2 K). (b), (c) $T$ dependence of the SC gap magnitude $\Delta(T)$ and the inverse QP lifetime $\Gamma(T)$ obtained by the fitting procedure using a Dynes function, respectively. The red curve in (b) represents a BCS curve with $2\Delta/k_B T_c$ = 3.9.

The coherence factors can be calculated from $\varepsilon_k$ and $E_k$ in the following way:

$$|u_k|^2 = 1 - |v_k|^2 = \frac{1}{2}\left(1 + \frac{\varepsilon_k}{E_k}\right). \quad (5)$$

On the other hand, the SC gap magnitude in the angle-integrated PES (AIPES) spectra is obtained through a fitting procedure using a Dynes function,[24] which represents the density of states (DOS) and is defined by

$$D(E, \Delta, \Gamma) = Re\left[\frac{E - i\Gamma}{\sqrt{(E - i\Gamma)^2 - \Delta^2}}\right]. \quad (6)$$

The FD function, including the sample $T$, is multiplied by the BCS spectral function or the Dynes function and further convoluted by the Gaussian function corresponding to the experimental energy resolution.

The red and gray open circles in Fig. 10(b) represent the AIPES spectra near $E_F$ for Sn taken below and above $T_c$, respectively.[20] A clear SC gap and a coherence peak were observed at 2.5 K. We estimated the SC gap magnitude $\Delta$ to be 520 μeV and the phenomenological broadening parameter (reciprocal of QP lifetime) $\Gamma$ to be 160 μeV through a fitting procedure using a Dynes function. The total energy resolution was estimated to be 400 μeV. Shown in Fig. 11(a) is the $T$ dependence of the PES spectra for Nb taken across $T_c$. The full gap and sharp coherence peak are well reproduced by the Dynes function, assuming isotropic SC gap symmetry with $(\Delta, \Gamma) = (1.5$ meV, 100 μeV$)$ at 2.6 K. With increasing $T$ toward $T_c$, the SC gap gradually closes and the spectrum finally follows a FD distribution [Fig. 11(a)]. By fitting SC spectra using the Dynes function, we can determine the $T$ dependence of $\Delta$ and $\Gamma$ as summarized in Figs. 11(b) and 11(c). $\Delta(T)$ clearly follows a BCS curve, and $\Gamma(T)$ has a nearly constant value at low $T$, with a slight increase toward $T_c$. A similar tendency was also observed for a $Pb_{0.9}Bi_{0.1}$ thin film by tunneling measurement, which was discussed in terms of the increase in the recombination process on approaching $T_c$.[24] These results suggest that the QP properties of low-$T_c$ superconductors can be investigated in detail by employing the second-generation laser-PES. In combination with an angle-resolving technique, laser-PES thus has significant capability to resolve the SC gap symmetry of low-$T_c$ superconductors. Some of these materials exhibit nodal (sign reversal) SC gaps indicative of unconventional electron pairing, as summarized in Figs. 12(a) and 12(b)[20,25–37]. For example, the non-centrosymmetric superconductor $LiPt_3B_2$ ($T_c = 2.8$ K) exhibits line nodes suggestive of parity mixing,[27] the HF superconductor $CeCoIn_5$ ($T_c = 2.3$ K) shows a signature of $d_{x2-y2}$ wave symmetry[28], and the Ru-oxide superconductor $Sr_2RuO_4$ ($T_c = 1.4$ K) exhibits spin-triplet superconductivity[38] with horizontal[33] or vertical line nodes.[34] Direct determination of the nodal position by laser-ARPES will contribute to the complete understanding of the exotic pairing mechanisms. The second-generation laser-PES system will lay a path to a new age of investigations on such low-$T_c$ superconductors.

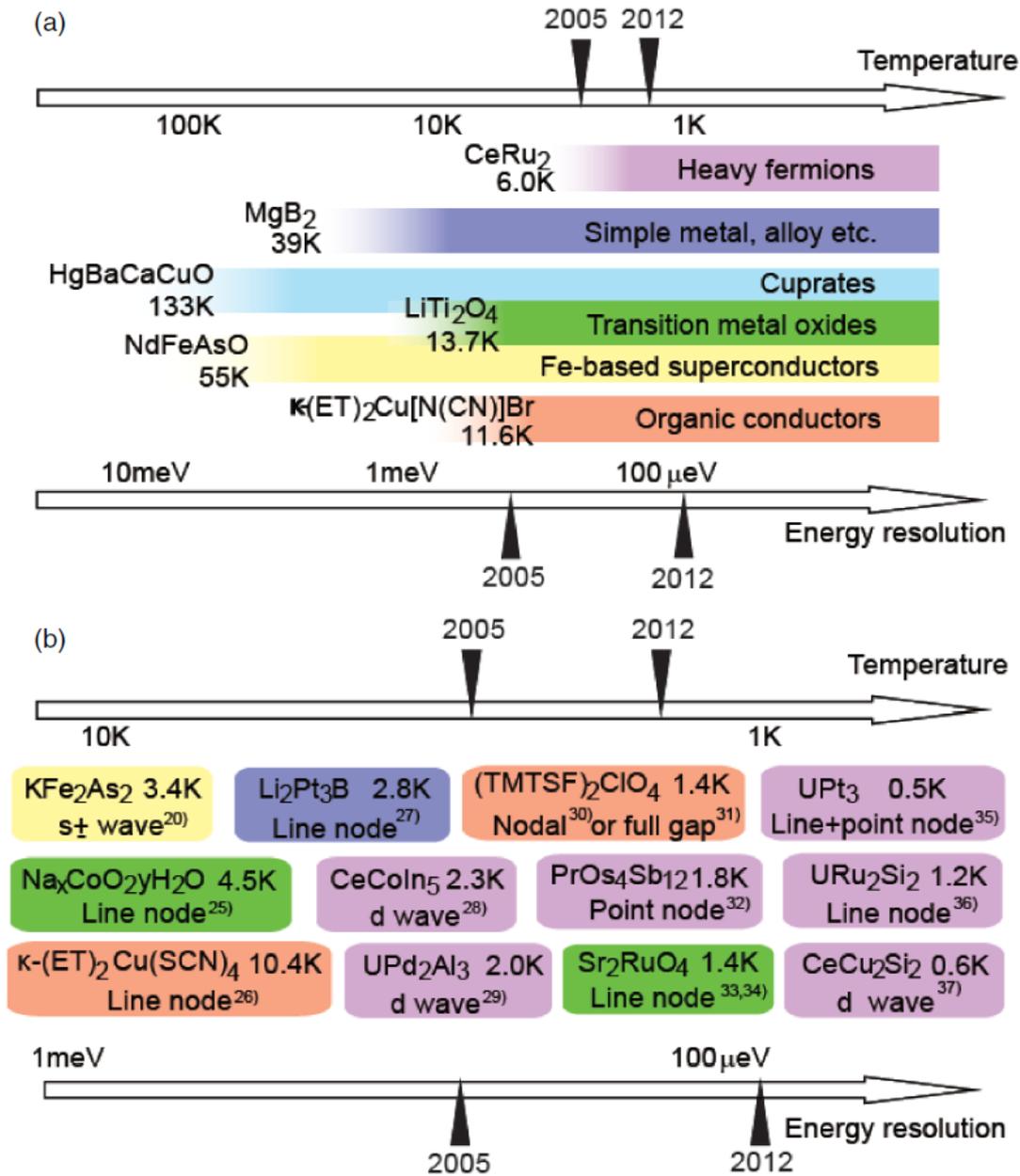

Fig. 12. (Color online) (a) Groups of superconductors with the highest $T_c$. (b) Unconventional low-$T_c$ superconductors and candidates for SC gap properties. The specifications of the first-generation and second-generation laser-PES systems are indicated by the black arrows labeled 2005 and 2012, respectively.

## 3. Applications of Laser Angle-Integrated Photoemission Spectroscopy

### 3.1 Investigation of the quasiparticle properties in the superconducting state

Low-$T$, high-energy-resolution laser-AIPES often exhibits a marked improvement in the sharpness of the spectra as compared with PES with other photon sources. Such an improvement allows us to directly investigate the QP properties in the SC state. Here we show examples of laser-AIPES results for the magnesium diboride system $Mg(B,C)_2$ ($T_{c\_max}$ = 39 K) and the pyrochlore oxide superconductor $KOs_2O_6$ ($T_c$ = 9.6 K).

#### 3.1.1 Two-gap superconductivity in $Mg(B,C)_2$

The superconductivity in $MgB_2$ has attracted much attention because $MgB_2$ has the highest $T_c$ among intermetallic superconductors.[39] Its $T_c$ of 39 K is close to the upper limit predicted by BCS theory.[40] The crystal structure of $MgB_2$ consists of alternating hexagonal layers of boron and magnesium. First-principles band calculations indicate that $MgB_2$ is a band metal with four bands crossing $E_F$.[41,42] Two of them are composed of the boron $2p$ $\sigma$ orbital with a two-dimensional character ($\sigma$ band), and the other two originate from the boron $2p$ $\pi$ orbital with a three-dimensional character ($\pi$ band).[43,44] It is generally accepted that $MgB_2$ exhibits multiple SC gaps,[45] originating from the two types of Fermi surface (FS) sheets with $\sigma$ and $\pi$ orbital characters.

In order to further understand the SC pairing mechanism of $MgB_2$, the perturbation of the electronic properties in the SC state induced by C substitution has been investigated by laser-AIPES.[46] Figure 13(a) shows the C substitution dependence of the SC gap features of $Mg(B,C)_2$. Owing to the high energy resolution of laser-AIPES, two peak features were clearly observed at $x = 0.0$ (red curve), while a single broad peak was observed for the PES with a He discharge lamp (black curve). These results are consistent with the ARPES result for $MgB_2$, reporting two-gap superconductivity on the $\sigma$ and $\pi$ bands.[43,44] Owing to the sharpness of the SC peaks, the SC gap magnitude was successfully estimated by a fitting procedure using two Dynes functions.[24] Figure 13(b) summarizes the $x$ dependence of the magnitude of each SC gap. The larger SC gap in the $\sigma$ band decreases with increasing $x$, while the smaller gap in the $\pi$ band is independent of $x$ up to $x = 0.075$. The $x$ dependence of $T_c$ shown by the solid curve in Fig. 13(b) nearly overlaps that of the larger gap. The comparable $x$ dependence between the larger SC gap and $T_c$ indicates that the value of $2\Delta/k_BT_c$, which represents the coupling strength of the Cooper pairing, is independent of $x$. At the weak-coupling limit, the value equals 3.52, known as the mean-field BCS value.[39] For the larger SC gap, $2\Delta/k_BT_c$ is estimated to be 4.1 and is nearly independent of $x$. These results suggest that the electron-phonon coupling in the $Mg(B,C)_2$ system is strong and that the superconductivity is mainly controlled by the electrons in the $\sigma$ band.

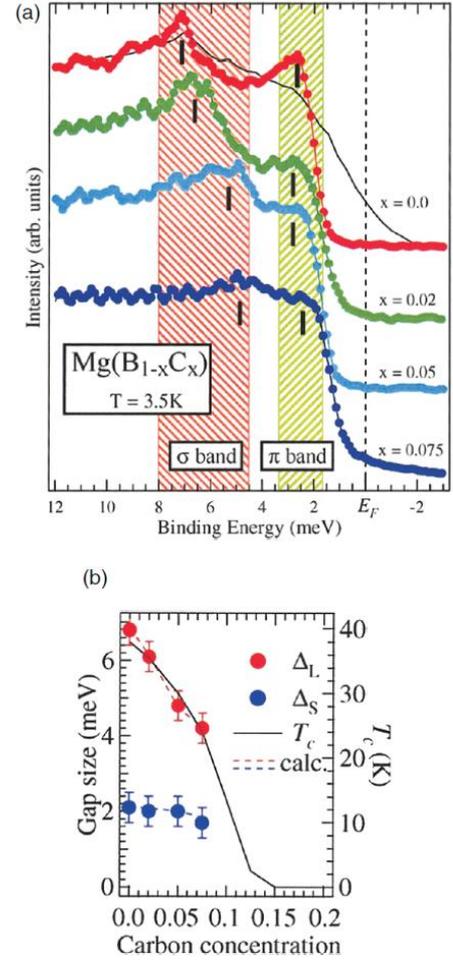

Fig. 13. (Color online) (a) AIPES spectra for $Mg(B_{1-x}C_x)_2$ polycrystalline samples with $x$ = 0.0, 0.02, 0.05, and 0.075 in the SC state ($T$ = 3.5 K) obtained by the VUV laser with energy resolution of 0.4 meV. The black curve indicates the PES spectrum for the polycrystalline sample at $x$ = 0.0 obtained by a helium discharge lamp with energy resolution of 3.0 meV. Vertical bars correspond to the SC gap size, and hatched areas emphasize the difference between the size of the two SC gaps. (b) $x$-dependence of the SC gap sizes (blue and red markers) and $T_c$ (solid curve). Reprinted from Ref. 46.

#### 3.1.2 Rattling and superconductivity in $KOs_2O_6$

A new series of $\beta$-pyrochlore superconductors $AOs_2O_6$ ($A$ = K, Rb, Cs) was discovered by Yonezawa and coworkers,[47–49] with relatively high $T_c$ of 9.6, 6.3, and 3.3 K, respectively. Experimental studies reported that $KOs_2O_6$ shows the most outstanding physical properties, such as the highest $T_c$, a high $H_{c2}$, strong electron-phonon coupling, and strong mass enhancement.[50–52] A number of anomalies have been discussed in terms of the rattling phenomenon. A K ion inside an $Os_{12}O_{18}$ cage moves 1 Å around its ideal site anharmonically because of the much smaller ion radius than the cage size. This is called rattling.[53] Specific heat measurements detected the anomaly at $T_p$ = 7.5 K (below $T_c$) only for $KOs_2O_6$; this is considered to be a first-order transition related to the rattling.[54] A theoretical study suggests that the independent rattling motion freezes at $T_p$, accompanied by the ordering of K ions.[55]

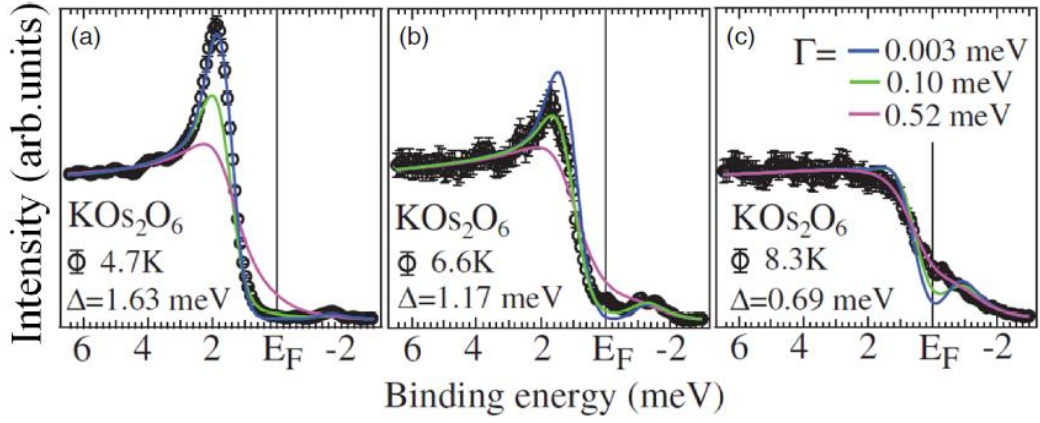

Fig. 14. (Color online) (a)–(c) Laser-AIPES spectra (black circles) for $KOs_2O_6$ taken at 4.7, 6.6, and 8.3 K, respectively. The solid curves represent the $s$-wave fitting functions with $\Gamma$ = 0.003, 0.10, and 0.52 meV, respectively, which are the values for the best fitting at each $T$. Reprinted from Ref. 56.

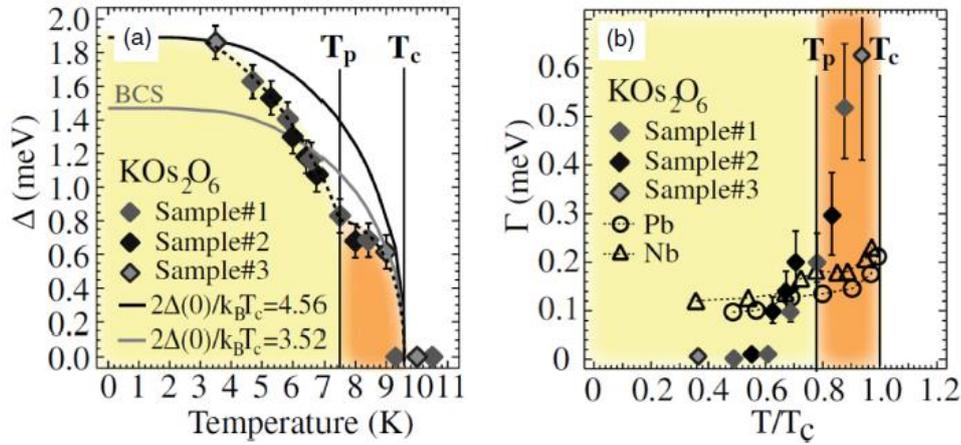

Fig. 15. (Color online) (a) $T$ dependence of $\Delta(T)$ for $KOs_2O_6$. The black broken curves are guides for the eyes. The gray and black solid curves represent the BCS curves with $2\Delta/k_BT_c$ = 3.52 (theoretical value for weak-coupling limit) and $2\Delta/k_BT_c$ = 4.56, respectively. (b) $T$ dependence of $\Gamma(T)$ for $KOs_2O_6$, Nb, and Pb. Reprinted from Ref. 56.

To investigate how the rattling phenomenon influences the QP properties in the SC state, laser-AIPES was employed for the examination of $KOs_2O_6$.[56] The black open circles in Figs. 14(a)–14(c) are the PES data for $KOs_2O_6$ at 4.7, 6.6, and 8.3 K, respectively. The blue, green, and pink curves represent fitting functions using the Dynes function.[24] Fitting functions with ($\Delta$, $\Gamma$) = (1.63 meV, 0.003 meV), (1.17 meV, 0.10 meV), and (0.69 meV, 0.52 meV) gave the best fit for 4.7, 6.6, and 8.3 K, respectively. The $T$ dependences of $\Delta$ and $\Gamma$ are summarized in Figs. 15(a) and 15(b), respectively. A noteworthy characteristic in the $T$ dependence of $\Delta(T)$ is the clear deviation from the BCS curve. The SC gap evolution seems to be suppressed between $T_c$ and $T_p$, then recovers rapidly below $T_p$. Such an anomaly at $T_p$ is also apparent in $\Gamma(T)$ as shown in Fig. 15(b). $\Gamma(T)$ versus $T/T_c$ is also plotted for several phonon-mediated superconductors, Pb ($T_c$ = 7.2 K, strong coupling, $2\Delta(0)/k_BT_c$ = 4.9) and Nb ($T_c$ = 9.2 K, weak coupling, $2\Delta(0)/k_BT_c$ = 3.8), which were similarly measured by the laser-AIPES. In strong contrast to Pb and Nb, $\Gamma(T)$ for $KOs_2O_6$ shows an unusually large increase at around $T_p$. The PES spectra at $T > T_p$ actually show extraordinary broadening compared with the sharp SC peaks observed at lower $T$. The rapid increase in $\Gamma(T)$ at around $T_p$, in great contrast to other superconductors, strongly indicates that the rattling phenomenon unique to $KOs_2O_6$ strongly affects the QP scattering at $T_p < T < T_c$. A similar effect has also been discussed for the normal-state resistivity realized under a magnetic field. It shows an unusually slow $T$ dependence for $T > T_p$, which eventually turns into $T^2$ behavior below $T_p$,[57] indicating that the electron scattering mechanism changes across $T_p$, from an anomalous scattering state at $T > T_p$ to a Fermi liquid (electron-electron scattering) state at $T < T_p$. A thermal conductivity measurement suggested that QP transport is dominated by electron-phonon scattering at $T_p < T < T_c$ and by electron-electron ($e$-$e$) scattering at $T < T_p$.[58] Laser-AIPES results seem to be consistent with these pictures. At the same time, the rattling motion might suppress the SC gap evolution at $T_p < T < T_c$. It is most likely that the rattling transition at $T_p$ markedly changes the QP scattering mechanism, also resulting in the unusual dip at $T_p$ in the SC gap evolution.

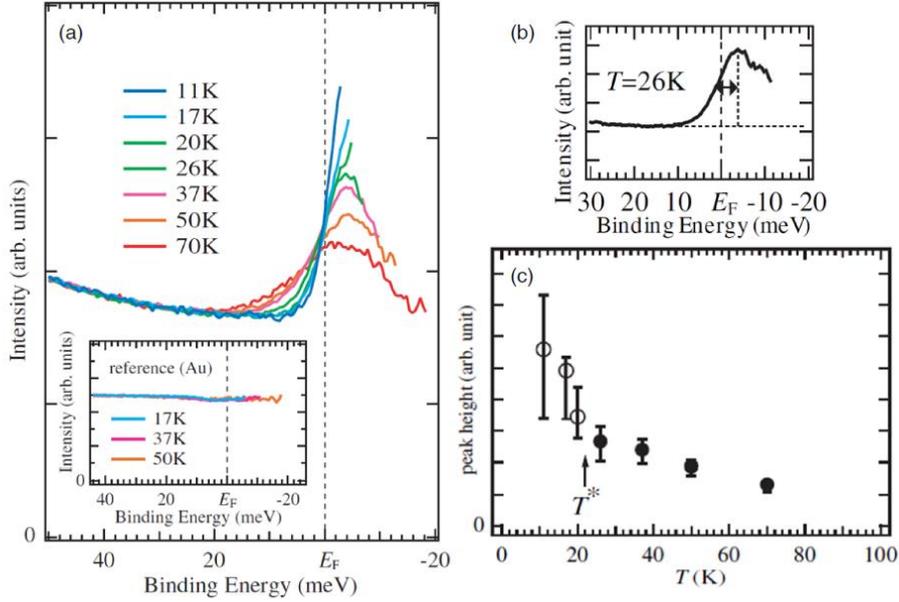

Fig. 16. (Color online) (a) $T$ dependence of the laser-AIPES spectra for $LiV_2O_4$ divided by the FD function including the sample $T$. The inset exhibits the spectra of gold as a reference, showing an absence of the peak feature near $E_F$. (b) Estimation of the peak height, peak position, and the background. (c) $T$ dependence of the peak height obtained by subtraction of the background intensity from the peak intensity. The filled circles show the peak height, while the open circles indicate the lower limit of the peak height below 26 K. $T^*$ corresponds to the crossover $T$. The error bar represents the experimental accuracy of $E_F$. Reprinted from Ref. 73.

*3.2 Observation of anomalous electronic structure near the Fermi level*

Anomalous physical properties are sometimes reflected in the electronic structure near $E_F$. Owing to the high energy resolution, unusual gap and/or peak evolutions near $E_F$ have been detected by laser-AIPES. The bulk sensitivity and high photon flux of the VUV laser also enable such investigation for chemically unstable materials and systems with low carrier density. In this section, we introduce three applications of laser-AIPES for the transition-metal oxide $LiV_2O_4$, the water-intercalated superconductor $Na_xCoO_2 \cdot yH_2O$, and B-doped diamond superconductors.

*3.2.1 Kondo resonance in 3d transition-metal oxide $LiV_2O_4$*

The discovery of HF-like behavior in $LiV_2O_4$ shed light on the possibility of a HF state solely composed of V 3$d$ electrons. The magnetic susceptibility and the electronic specific heat coefficient ($\gamma$) show significantly large values below $T^* \approx 20$ K.[59] $\gamma$ is obtained to be ~420 mJ/(molK$^2$), which is about 25 times larger than that estimated from the bare mass obtained by local-density-approximation (LDA) band calculations.[60] This value is close to that of typical $f$-electron HF compounds.[61] The electrical resistivity $\rho$ in $LiV_2O_4$, on the other hand, exhibits $T^2$ dependence, $\rho = \rho_0 + AT^2$, with an enhanced $A$ at $T \ll T^*$, which is another property of the HF state.[62] The LDA calculation for $LiV_2O_4$ indicates that the three bands derived from the V 3$d$ orbitals cross $E_F$ (a doubly degenerate $E_g$ and a nondegenerate $A_{1g}$ with bandwidths of 2 and 1 eV, respectively.)[60] By assuming $A_{1g}$ as the "localized band" and $E_g$ as the "itinerant bands" in analogy with the case of $f$-electron compounds, the possibility of a HF state being created by the Kondo effect was suggested for the first time in a $d$-electron system.[63] However, since their band widths are not so different, it has been discussed whether the HF-like state in $LiV_2O_4$ can be explained within a simple Kondo scenario. Another important characteristic of $LiV_2O_4$ is the cubic spinel structure that gives rise to the geometrical spin frustration.[64] $LiV_2O_4$ shows Curie-Weiss-like behavior in the magnetic susceptibility in the $T \gg T^*$ region, indicative of the existence of local magnetic moments.[59] However, no evidence of long-range magnetic order (or a spin glass) has been reported down to 0.02 K,[59] possibly due to suppression by the geometrical magnetic frustration. In addition, recent neutron scattering[65,66] and muon spin relaxation[67,68] experiments indicate that local magnetic moments with short-range correlation exist down to $T \ll T^*$. The possible role of dynamical spin fluctuation has thus often been discussed in an attempt to understand the anomalous thermal, magnetic, and transport properties in $LiV_2O_4$.[69–71]

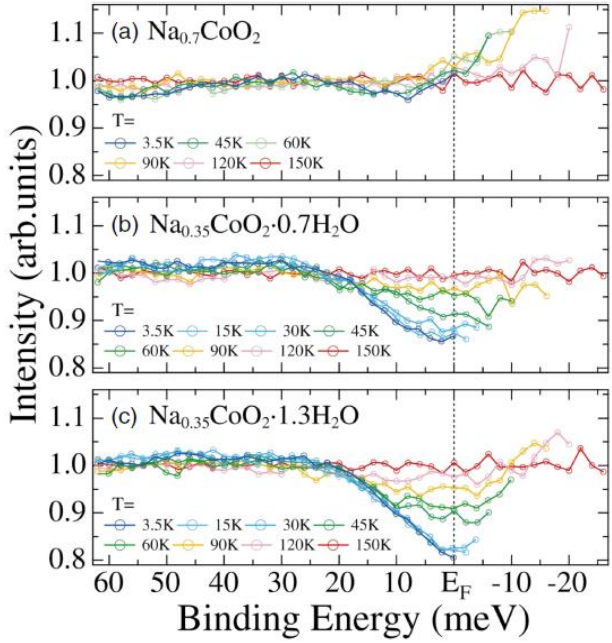

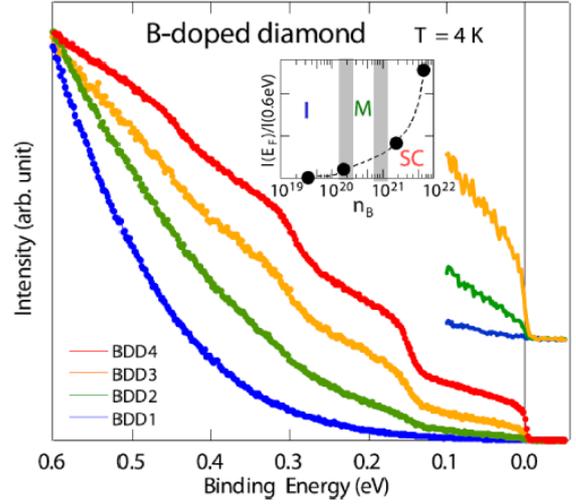

Fig. 17. (Color online) (a)–(c) $T$ dependence of the normalized laser-AIPES spectra divided by FD function for NCO, NCO07, and NCO13, respectively. Reprinted from Ref. 90.

Fig. 18. (Color online) Doping dependence of the laser-AIPES spectra for B-doped diamond (BDD) superconductors. The near-$E_F$ region for BDD1–BDD3 is magnified in the upper part. The inset shows the intensity at $E_F$ normalized at 0.6 eV as a function of $n_B$. I, M, and SC stand for the insulator, metal, and superconducting phases as the ground state, respectively. Reprinted from Ref. 105.

To clarify the origin of the HF-like state in LiV$_2$O$_4$, investigation of the low-energy electronic structure around $E_F$ is necessary. For example, the observation of a Kondo resonance (KR) peak near $E_F$ in CeCu$_2$Si$_2$ revealed the usefulness of high-resolution PES for this purpose.[72] Using the advantageous features of laser-PES, a sharp peak located above $E_F$ was observed in the PES spectrum divided by the FD function [Fig. 16(a)].[73] The height and position of the peak were estimated in the manner indicated in Fig. 16(b). The $T$ dependence of the peak height in Fig. 16(c) shows a steep increase below ~20 K and was found to be comparable to the KR observed in conventional $f$-electron HF compounds. In a Kondo scenario, evolution of the KR peak indicates that the mass renormalization becomes rapidly enhanced below $T_K$ due to the strong hybridization between localized and itinerant electrons. From the results of calculations based on the two-band Hubbard model,[74] it has been proposed that the QP state with a sharp KR peak can be similarly formed in LiV$_2$O$_4$, even though the widths of the $A_{1g}$ and $E_g$ bands are not in extreme contrast. In the calculations, the interband Coulomb repulsion plays a crucial role by strongly emphasizing the small difference between the two bands, making one of the $d$ electrons localized in the lower $A_{1g}$ band. As a result, strong renormalization due to the Kondo effect occurs, which can create a sharp peak just above $E_F$ in the DOS. The peak in the PES spectra is similar to the KR peak obtained in this model. Thus, such a Kondo scenario might be an appropriate candidate for explaining the HF-like behavior in LiV$_2$O$_4$.

3.2.2 *Pseudogap in water-intercalated superconductor Na$_x$CoO$_2 \cdot y$H$_2$O*

Following the discovery of the hydrated cobalt oxide superconductor Na$_{0.35}$CoO$_2 \cdot$1.3H$_2$O (NCO13) ($T_c$ = 4.5 K),[75] extensive experimental and theoretical studies have been carried out. This is because the conductive two-dimensional CoO$_2$ layers can be regarded as an electron-doped correlated $S$ = 1/2 triangular network of frustrated Co spins,[75] where superconductivity emerging from a non-Fermi-liquid ground state has been proposed.[76] Theoretical studies have proposed a variety of SC gap symmetries depending on the model used.[77–80] While several experimental studies have been performed, there is no general consensus regarding its pairing symmetry.[81–84] In the cuprates, whose electronic structure is often described in terms of a quasi-two-dimensional correlated system, investigations of the unusual normal-state physical properties have led to a deeper understanding of the unconventional superconductivity. Even among the layered Co oxides, a rich phase diagram has been proposed for the non-hydrated Na$_x$CoO$_2$ system.[85] For hydrated SC samples, as has also been suggested from theoretical studies,[86,87] important roles of the well-separated Co$^{3+}$ and Co$^{4+}$ in the SC Co oxides have been reported following the observation of Co 2$p$ core-level peaks using hard X-ray PES.[88] The coexistence of superconductivity and other ordered states has also been discussed in cuprates.[89] In this sense, NCO13 provides another opportunity to investigate charge and/or spin order competing or cooperating with superconductivity in these oxides.

In order to reveal the normal-state electronic properties of the SC Co oxides, laser-PES was employed for NCO13 and related materials.[90] PES under an ultrahigh vacuum requires special caution because SC NCO13 tends to become non-SC Na$_{0.35}$CoO$_2 \cdot$0.7H$_2$O (NCO07) due to the loss of H$_2$O molecules. Hydrated samples were therefore carefully treated as follows. First, the polycrystalline samples were covered with silver paste and mounted on copper substrates to prevent the loss of water molecules under vacuum. Then, the prepared samples were cooled to 180 K and fractured *in situ*. Immediately after the fracturing, they were transferred to a measurement chamber and measured without being warmed to above 180 K. After these sample treatments, the DOS near $E_F$ was successfully obtained for the SC NCO13, the NSC NCO07, and the parent $Na_{0.7}CoO_2$ (NCO) by employing laser-AIPES.[90] As summarized in Figs. 17(a)–17(c), pseudogap (PG) formation with an energy scale of 20 meV was observed for NCO13 and NCO07, but was clearly absent for NCO. The further evolution of a smaller PG of 10 meV below 45 K only detected in NCO13 highlights a difference from NCO07, indicating the validity of the above-mentioned sample treatment. The energy scale of the PG is much larger than the expected magnitude of the SC gap, suggestive of an additional competing order parameter at low $T$ in NCO13 and NCO07. Owing to not only the bulk sensitivity but also the high photon flux of laser-AIPES, we were able to detect the anomalous normal-state electronic properties of the water-intercalated cobalt oxide superconductor.

*3.2.3 Multiphonon sidebands in B-doped diamond superconductors*

Diamond is known to be an excellent electrical insulator and also a good thermal conductor. The thermal conductivity originates from its $sp^3$ bonding, leading to high-energy optical phonons (HOPs) and a Debye $T$ of 1860 K. Boron (B) doping induces high-mobility p-type shallow acceptor states at 0.37 eV in the band gap. On increasing the B concentration ($n_B$), the system undergoes an insulator–metal transition at $n_B^{MI} \approx 2 \times 10^{20}$ cm$^{-3}$ (~0.1 % substitution), followed by the emergence of superconductivity (inset of Fig. 18).[91,92] The superconductivity in B-doped diamond (BDD) is believed to be of the conventional BCS type, with HOPs as the possible pairing glue.[93–95] However, the coupling of dilute carriers with such HOPs of ~150 meV is unique, raising questions about the validity of the adiabatic approximation[96,97].

In the photoemission process, an incident photon is absorbed by a solid and an electron is then removed, *i.e.*, an $N$-electron system becomes an ($N-1$)-electron system. In the sudden approximation, and for sufficiently weak interactions allowing a one-particle description of the $N$ electrons, the PES spectrum shows $\delta$-function-like peaks for each initial state of the $N$-electron system. In contrast, for a strongly interacting system, several excited states can contribute to the $N-1$ final states, and lead to a multiple-peak spectrum for the initial state.[98,99] When the electrons couple with a single boson mode, the variation of the intensities in the multiple peaks is known to be given by the Franck-Condon factor (a Poisson distribution). Such a characteristic spectral line shape can be reproduced with a model based on the localized electrons that couple to a harmonic oscillator.[100] The broad spectral shapes of the valence band in some transition-metal compounds exhibiting a charge order, as well as layered high-$T_c$ cuprates,[101–104] were explained by the envelope of multiphonon sidebands arising from strong e-ph coupling. However, clear evidence of phonon sidebands in the valence band has not yet been detected by PES.

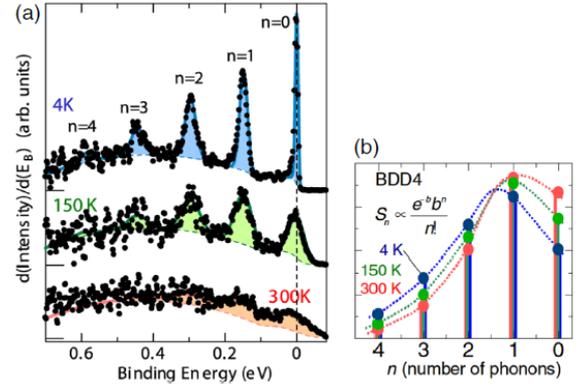

Fig. 19. (Color online) (a) $T$ dependence of the $E_B$-derivative laser-AIPES spectra. The solid (broken) curves represent the fitting function (background) for each $T$. (b) $T$ dependence of sideband intensities $S_n$ obtained by the Franck-Condon analysis of BDD4. The broken curves are guides for the eyes to show the envelope of the Poisson distribution. Reprinted from Ref. 105.

To elucidate the coupling of dilute carriers with HOPs, laser-AIPES was employed to observe the valence band for BDD1 (insulating, $n_B = 3.5 \times 10^{19}$ cm$^{-3}$), BDD2 (weakly insulating, $n_B = 1.75 \times 10^{20}$ cm$^{-3}$), BDD3 (SC, $n_B = 1.88 \times 10^{21}$ cm$^{-3}$), and BDD4 (SC, $n_B = 6.53 \times 10^{21}$ cm$^{-3}$).[105] Figure 18 shows the $n_B$ dependence of the laser-AIPES spectra obtained at 4 K for BDD1–BDD4. While the BDD1 sample exhibits a long-tail-like spectrum without any evidence of a Fermi edge, indicative of a disordered p-type semiconductor, the weakly insulating BDD2 shows a small Fermi edge. The evolution of the Fermi edge through B doping is apparent in the magnified spectra plotted in Fig. 18. A rapid increase in the carrier number across the insulator–metal–superconductor transitions is evident, thus confirming degenerate metal behavior for samples with $n_B > n_B^{MI}$. For the metallic samples, anomalous steplike features appear at around $E_B = 150, 300,$ and 450 meV in addition to the Fermi edge. Periodic energy steps of $150 \times n$ meV ($n = 1, 2, 3, \cdots$) are clear, and these are most prominent for the BDD4 sample. Such periodic structures are possibly multiphonon sidebands emerging as replicas of the Fermi edge due to strong e-ph coupling. The optical phonon branches in pristine diamond are well known, and they correspond to the bond-stretching vibration mode that is triply degenerate at the $\Gamma$ point with an energy of 164 meV.[106] An infrared reflectivity measurement of a SC BDD also shows a clear kinklike structure at ~150 meV.[95] This suggests that the periodic steps in laser-AIPES spectra are due to strong coupling of the optical phonons ($\omega_{ph} \approx 150$ meV) to the doped carriers, resulting in the multiphonon sidebands.

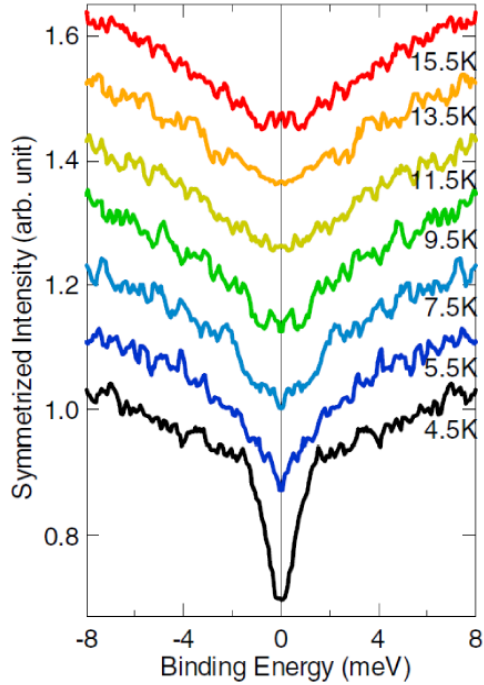

Fig. 20. (Color online) $T$ dependence of the laser-AIPES spectrum for B-doped diamond (BDD) ($T_c$ = 6.6 K) symmetrized at $E_F$. Reprinted from Ref. 109.

The $T$ dependence of the periodic steps for BDD4 was further investigated. In the $E_B$-derivative PES spectra shown in Fig. 19(a), the steplike features now appear as well-defined peak structures. The sidebands become unclear at higher $T$ as the Fermi edge is smeared owing to the FD distribution. To further investigate the multiphonon sidebands, Franck-Condon analysis was applied. The $E_B$-derivative spectrum at 4 K was fitted with a background function $I_{BG}(E_B)$, represented by the broken curve, and a set of normalized Gaussian curves $g(E_B - n\omega_0, W_n)$ with peaks at $E_B = n\omega_0$ and half width at half maximum (HWHM) $W_n$, written as $I(E_B) = I_{BG}(E_B) + \Sigma S_n g(E_B - n\omega_0, W_n)$. The relation $W_n = (W_0^2 + nW_{ph})^{1/2}$ was assumed.[105] The integrated intensity of the $n$th peak $S_n$ was fixed to follow the Poisson distribution $S_n = e^{-b}b^n/n!$. The fitting result for the data at 4 K is shown by the blue curve in Fig. 19(a), with the best fitting parameters being $b$ = 1.6, $\omega_0$ = 148 meV, $W_0$ = 4.8 meV, and $W_{ph}$ = 12 meV. The zero-phonon line corresponds to the Fermi-edge derived peak at $E_F$. Here, the factor $b$ is the effective $e$-ph coupling parameter. The obtained $S_n$ are indicated as a function of $n$ and $T$ in Fig. 19(b).

Since $b$ = 1.6 (> 1), $S_1$ is greater than $S_0$ at 4 K, indicative of the suppression of the zero-phonon line and the phonon dressing of carriers. The applicability of Franck-Condon analysis to the multiphonon sidebands in BDD4 clearly suggests that they are created in an optical excitation process and indicates that the adiabatic approximation is valid for the coupling of carriers with HOPs. On increasing $T$, $b$ tends to become smaller, indicative of the spectral weight transfer from higher- to lower-$E_B$ sidebands. In other words, the $e$-ph coupling constant $b$ systematically increases with decreasing $T$. Since the characteristics and population of HOPs are not expected to change significantly up to 300 K, such $T$ dependence must be electronic in origin. It is noteworthy that a simple localized $e$-ph model is applicable to carriers in a metal undergoing a SC transition. One possible way to deal with such localized carriers is the small polaron picture, previously discussed for the broad PES line shape in transition-metal compounds.[101–104] Theoretical calculations for polaronic systems have predicted periodic oscillations in the PES spectrum near $E_F$.[107,108] This seems a plausible picture for metallic diamond, and may explain this first observation of phonon-derived oscillations in near-$E_F$ PES spectra. This unique nature of the carrier–phonon coupling might lead to the superconductivity in doped diamond.

Laser-AIPES further revealed a SC gap opening at $E_F$ for a BDD thin film with $n_B$ = 8.4 × 10²¹ cm⁻³.[109] Although the symmetrized spectrum lacks a QP peak (Fig. 20), the dominant size of the SC gap is estimated to be 0.78 meV at 4.5 K, indicative of weak-coupling superconductivity. As indicated by a dip near $E_F$, the SC gap was observed up to 11 K, while $T_c$ of the magnetization was ~6.6 K. Such a deviation might reflect the inhomogeneous conductivity near $T_c$, suggesting its high sensitivity when probing the local SC state with high $T_c$.

## 4. Applications of Laser Angle-Resolved Photoemission Spectroscopy

*4.1 Observation of the band dispersions near the Fermi level*

ARPES of the near-$E_F$ electronic structure provides rich information, including the shapes of the FSs, and the effective mass and velocities of the carriers. Anomalous kinks are sometimes observed in the band dispersions below $E_F$. These kinks originate from a coupling with the collective modes. Investigation of the kinks might lead to an understanding of the nature of the collective modes and the pairing mechanism of the unconventional superconductivity. In such studies, the high energy and momentum resolution of laser-ARPES can be strong advantages. In this section, we introduce two examples of the investigations of the low-energy kinks for the high-$T_c$ cuprate $(Bi,Pb)_2(Sr,La)_2CuO_{6+\delta}$ and topological insulator $Bi_2Se_3$ families.

*4.1.1 Low-energy band renormalization in $(Bi,Pb)_2(Sr,La)_2CuO_{6+\delta}$*

A major issue in researching the high-$T_c$ SC mechanism is to identify the bosons that participate in the electron pairing. An electron coupling with a bosonic mode modifies the band dispersion within its excitation energy. The band renormalization kink has been detected in some cuprates and is attracting considerable attention because of the prospect that the collective mode plays an essential role in the electron pairing. However, the origin of the kink, particularly whether it originates from phonons or spin excitations, still remains controversial, mostly because of the existence of different modes at almost the same energy.[110–112] Previous ARPES studies uncovered a noteworthy band renormalization near $E_F$ (< 20 meV),[113–115] in addition to the well-studied kinks appearing at 40–80 meV, in $Bi_2Sr_2CaCu_2O_{8+\delta}$ (Bi2212) ($T_{c\_max}$ = 95 K). A recent theoretical study suggested that the coupling to phonons is too small to account for the significant band renormalization observed in cuprates.[116] Other studies pointed out that strong electron correlation or a reduced screening effect can significantly enhance the electron-phonon coupling.[115,117–120]

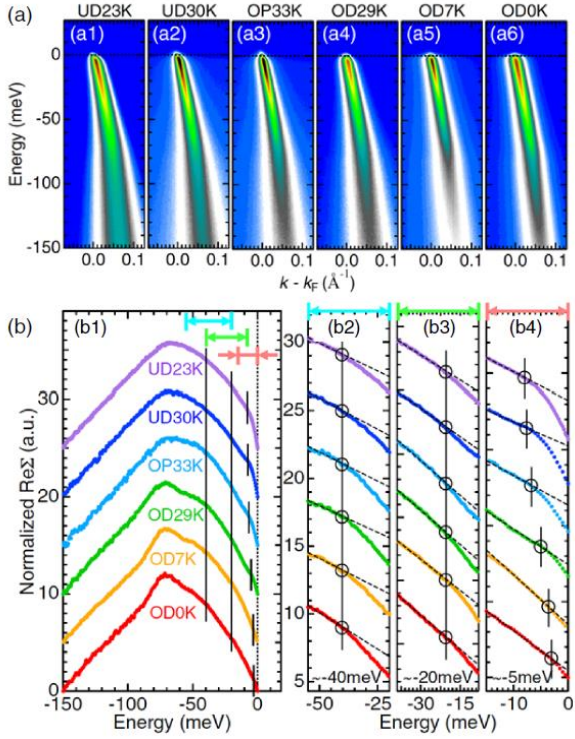

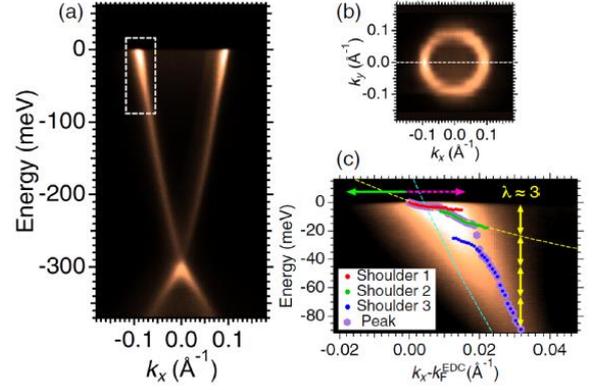

Fig. 21. (Color online) (a1)–(a6) Doping dependence of the nodal $E$-$k$ images for Bi2201 from underdoped to overdoped samples. (b1) Doping dependence of normalized Re $\Sigma$. Black bars indicate the energy positions of the kinks. (b2)–(b4) Spectra in (b1) with an enlarged energy scale [arrows in (b1)]. Kink positions are indicated with circles and bars. Reprinted from Ref. 121.

In order to investigate the low-energy kinks in a wide doping range, the $(Bi,Pb)_2(Sr,La)_2CuO_{6+\delta}$ (Bi2201) ($T_{c\_max}$ = 35 K) system was studied by laser-ARPES.[121] The doping dependence of the band dispersion is shown clearly in Figs. 21(a1)–21(a6). The kinks were detected at ~40, ~20, and ~3 meV for an overdoped sample. Binding energies that were almost doping-independent in the multiple mode couplings were found at values of ~20 and ~40 meV. In contrast, the energy scale and coupling constant of the lowest-energy kink increase toward the underdoped region, with an abrupt enhancement across the optimal doping [Figs. 21(b1)–21(b4)], suggesting that electron correlation is an important factor in the development of low-energy band renormalization. The doping dependence strongly contrasts with the nature of the well-known kink at 70 meV, which is common to different cuprate families. These observations highlight a different origin of the lowest-energy kink from the others. The nontrivial behavior along the node supports the theoretical idea that forward scattering arising from the interplay between the electrons and in-plane polarized acoustic phonon branch is the origin of the low-energy band renormalization. This finding is crucial because it has been intensively discussed whether or not phonons can produce the strong renormalization observed in cuprates.[116,120] A comparison of the low-energy kinks between Bi2201 and Bi2212, with $T_c$ 2.5 times higher than that of Bi2201, further confirms the universality of the low-energy kink in cuprates, implying a close relationship between the low-energy renormalization and $T_c$ or the size of the energy gap close to the node.[121]

Fig. 22. (Color online) (a) $E$-$k$ image of $Bi_2Se_3$ taken along $\Gamma$–$K$ line [dashed line in (b)]. (b) FS profile. (c) $E$-$k$ image within a narrow range near $E_F$ marked by the broken rectangle in (a); parabolic bands (dashed curves) with masses of 0.83 $m_e$ and 0.14 $m_e$[133] are superimposed. Reprinted from Ref. 132.

### 4.1.2 Topological surface states of $Bi_2Se_3$ and $Cu_xBi_2Se_3$

Topological insulators (TIs) have surface states dominated by massless Dirac fermions that are topologically protected from disorder scattering.[122] It has been elucidated that the Dirac band dispersion is anomalously renormalized by effects such as $e$-$ph$ interaction, electron-hole pair generation, and electron-plasmon coupling, leading to various intriguing physical properties.[123–125] While TIs are essentially understood within the non-interacting topological theory,[126–128] Dirac fermions in actual materials may be influenced by nontrivial many-body interactions. Since the Dirac fermions in TIs are distinguished from those in graphitic materials in terms of their helical spin texture as well as possible interactions with a bulk electronic state, the low-energy excitations in the topological surface state are of particular interest. For example, a pronounced Kohn anomaly indicating a strong $e$-$ph$ coupling was observed in the surface phonon branch of $Bi_2Se_3$;[129] scanning tunneling spectroscopy (STS) uncovered an intriguing feature with sharp peaks at low energies (< 20 meV) in the Landau-level spectra,[130] indicating an anomalous increase in the QP lifetime near $E_F$. It has been theoretically proposed that a novel low-energy collective mode called a spin plasmon emerges as a consequence of the spin-momentum locking in the topological surface state.[131]

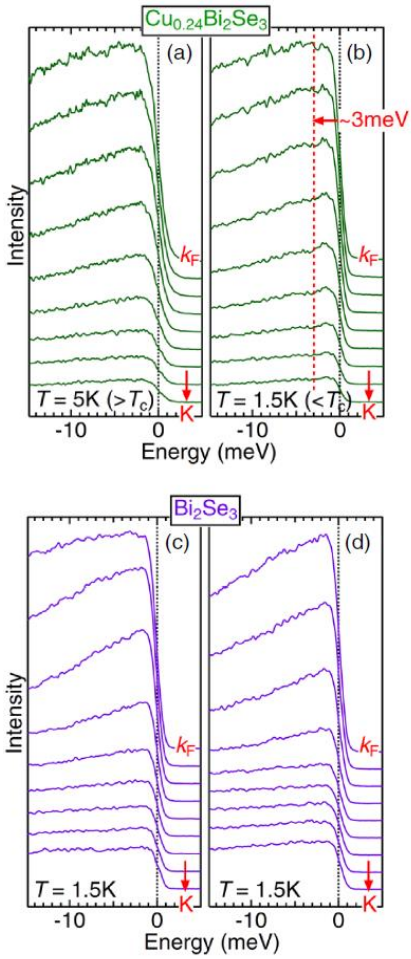

Fig. 23. (Color online) (a), (b) EDCs of $Cu_{0.24}Bi_2Se_3$ close to $k_F$ measured above and below $T_c$, respectively. The dashed line in (b) indicates the spectral dip feature. (c), (d) EDCs of pristine $Bi_2Se_3$ close to $k_F$ for aged and fresh surfaces, respectively. Reprinted from Ref. 132.

To elucidate the Dirac dispersion of the topological surface state near $E_F$, the second-generation laser-ARPES was employed to investigate $Bi_2Se_3$ and $Cu_xBi_2Se_3$.[132] As shown in Fig. 22(a), a sharp Dirac dispersion was observed for $Bi_2Se_3$ along the $\Gamma$–$K$ line, indicated by a dotted line in the FS profile in Fig. 22(b). Using enlarged energy and momentum scales, strong mode couplings at binding energies of ~15–20 and ~3 meV were resolved as indicated in Fig. 22(c).[133] The coupling to the $A_{1g}^2$ phonons is proposed as the candidate for the former mode. For the latter mode, there are two possible origins: one is the optical mode of surface phonons and the other is the spin plasmons, which have been proposed to be the low-energy excitations of helically spin-polarized Dirac fermions. Despite the extremely large mass enhancement factor of 3, the topological surface state does not exhibit any band reconstruction down to the lowest $T$, indicating that the helical Dirac cone is protected from the density-wave formations naturally expected for a system with extremely strong couplings to bosons.

Intriguingly, a peak-dip-hump structure in the energy distribution curves (EDCs), which is often considered as a signature of the kink in the dispersion, becomes visible at 3 meV in SC samples of Cu-doped $Bi_2Se_3$. Figures 23(a) and 23(b) show the EDCs of $Cu_{0.24}Bi_2Se_3$ ($T_c$ = 3.5 K) measured below and above $T_c$, respectively. The peak-dip-hump structure can be seen at 3 meV below $T_c$, while it is absent above $T_c$. Pristine $Bi_2Se_3$ was also examined under the same condition ($T$ = 1.5 K), but it did not show the peak-dip-hump structure [Fig. 23(d)]. The absence of the peak-dip-hump structure is also confirmed for an aged surface of $Bi_2Se_3$ [Fig. 23(c)]. Obviously, the enhancement of the 3 meV mode coupling has some relation to the superconductivity, and the origin of this enhancement needs to be further investigated in the future.

### 4.2 Investigation of the superconducting mechanism of iron-based superconductors

The iron-based superconductors[134–136] are the second class of high-$T_c$ materials after the cuprates. Superconductivity emerges via the application of pressure or carrier doping into the parent materials, which undergo a structural transition wherein a high-$T$ paramagnetic (PM) metal becomes a low-$T$ antiferromagnetic (AF) metal (Fig. 24).[137–139] While the antiferromagnetism of cuprates is derived from a localized Mott insulator, its role in the superconductivity in iron-based compounds is intriguing,[140] particularly because the high-$T$ undoped phase is an itinerant metal.[141–143] Possible roles of the multiband character in terms of the five Fe 3$d$ orbitals forming multiple FSs should also be considered seriously, in contrast to single-band cuprates.

Laser-ARPES is suitable for research into the iron-based superconductors because of the following advantages. The first is the variety of light polarizations, which allows orbital-selective observations of the band dispersions near $E_F$. The second is the high energy resolution and cooling ability, which are useful for detecting the precise anisotropy of the SC gap symmetry. Some low-$T_c$ iron-based superconductors, such as $KFe_2As_2$ (K122), are expected to show nodal SC gaps.

#### 4.2.1 Ground state of the parent compound $BaFe_2As_2$

The parent compound $BaFe_2As_2$ (Ba122) exhibits a stripe-type AF ordering at $T_N$ accompanying a structural transition from a tetragonal to orthorhombic lattice at $T_s$ [Figs. 25(a) and 25(b)].[144] ARPES measurements on the Ba122 family reported three hole and two electron FS sheets at the Brillouin zone (BZ) center and corners (insets of Fig. 24), respectively, which are composed of multiple Fe 3$d$ orbitals.[145,146] Considering the nesting tendency connecting quasi-cylindrical FSs, electron pairing mediated by AF spin fluctuations has been proposed.[147,148] Hence, the elucidation of the electronic structure of the parent materials, especially in the AF state, is essential to reveal the unconventional superconductivity in the iron-based superconductors.

However, it has been difficult to discuss the impact of the magnetostructural transition on the electronic structure of the parent compounds owing to the mixing of twinned orthorhombic domains.[149–151] Since a typical domain size is very small (~1 μm), the electronic structures observed by macroscopic probes will inevitably overlap with those rotated by 90°. s- and p-polarized VUV lasers have been employed in order to detect the electronic structure in an orthorhombic single domain of the parent compound Ba122.[152] While the size of the laser spot (~200 μm) is much larger than that of the orthorhombic domains [Fig. 26(a)], the electronic structure from each of the twinned domains was successfully extracted owing to matrix element effects that make the ARPES intensity strongly dependent on the light polarization and sample geometry.

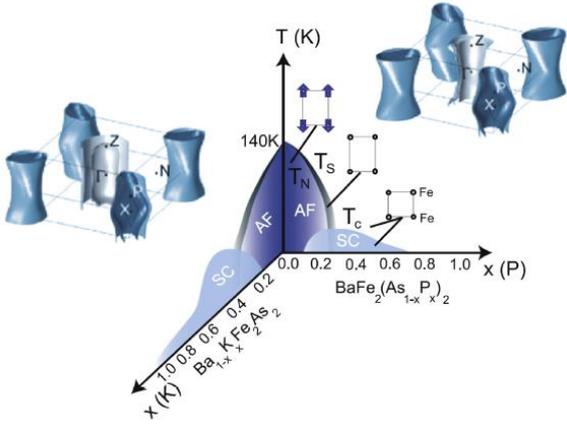

Fig. 24. (Color online) Schematic phase diagram of $BaFe_2(As,P)_2$ and $(Ba,K)Fe_2As_2$. $T_N$ represents the AF transition temperature and $T_s$ represents the tetragonal-to-orthorhombic structural transition temperature. Blue arrows indicate stripe-type AF spin ordering on the orthorhombic lattice. Insets show calculated FSs of the optimally doped composition for both systems. Reprinted from Ref. 166.

Figures 25(c) and 25(d) show the FS profiles measured by a p-polarized laser above and below $T_N$, respectively.[152] Above $T_N$, the FS shows a nearly circular shape with fourfold symmetry. In contrast, the FS below $T_N$ splits into a pair of large FSs along the orthorhombic axis $a^o$ and a pair of small FSs along axis $b^o$, thus resulting in significantly modified FSs with twofold symmetry. Furthermore, when the linear polarization is changed from p to s at the same sample position, the twofold FSs in the low-T AF state rotate by 90° [Figs. 25(d) and 25(f)], while very little change is observed above $T_N$ [Figs. 25(c) and 25(e)]. Considering the selection rules,[3] the complete 90° rotation of the FSs implies that the FS of each single domain has a particular orbital character and is alternatively active to s- or p-polarization. In the experimental configuration used, with the incidence and emission planes corresponding to the mirror planes of the crystal [Figs. 26(b) and 26(c)], the initial states having even (odd) parity with respect to the mirror plane are active to p (s)- polarization.[3] The detectable d orbitals following this selection rule are summarized for each possible relation between the polarization and orthorhombic orientations, as depicted in Figs. 26(b) and 26(c). It is notable that the orthorhombic orientation in domain A is rotated 90° with respect to that in domain B, and that the permissible combination of experimental geometries is limited to case 1 or case 2. Thus, we can observe the 90° rotation of FSs only if the FS is composed of a single $d_{zx}$ orbital (zx) for case 1 or a single $d_{yz}$ orbital ($yz$) for case 2 ($x$, $y$, and $z$ are the coordinates along the crystal axes of the orthorhombic settings $a^o$, $b^o$, and $c^o$, respectively).

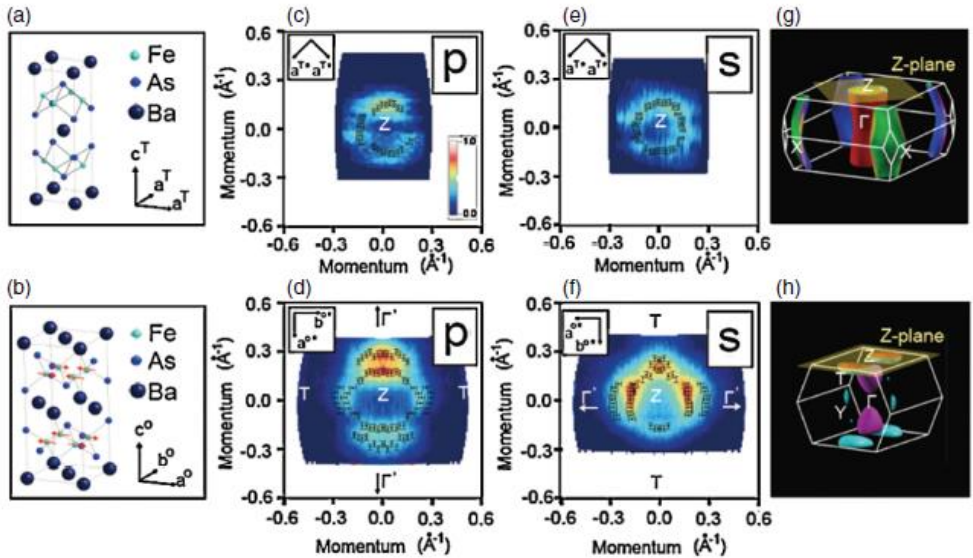

Fig. 25. (Color online) (a) Crystal structure of Ba122 in the tetragonal structure. (b) Crystal and magnetic structure in the stripe-type AF ordered orthorhombic structure. (c), (d) FS profile of Ba122 measured by p-polarization at 180 K (above $T_N$) and 30 K (below $T_N$), respectively. (e), (f) FS profile of Ba122 measured by s-polarization at 180 K (above $T_N$) and 30 K (below $T_N$), respectively. (g), (h) Whole FSs in the first BZ obtained by local density approximation calculations considering paramagnetic tetragonal and stripe-type AF orthorhombic structure, respectively, using the experimentally obtained structural parameters. Reprinted from Ref. 152.

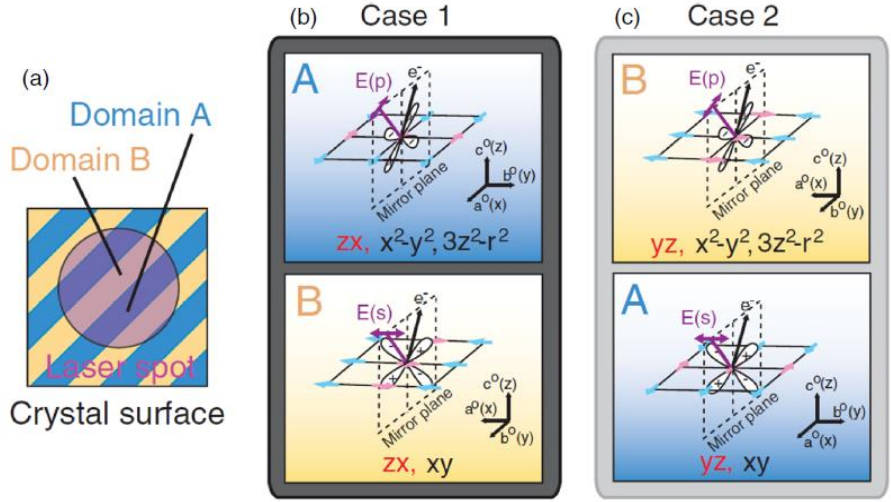

Fig. 26. (Color online) (a) Schematic of orthorhombic twinned structure at $T < T_s$. The orthorhombic orientation in domain A is rotated 90° with respect to that in domain B. (b), (c) Two permissible cases of the relation between the polarization and orthorhombic orientations, denoted as case 1 and case 2. Arrows on the square lattice represent the stripe-ordered Fe spins. Reprinted from Ref. 152.

First-principles LDA band calculations show FS profiles in the PM and AF states that are comparable to those obtained by laser-ARPES [Figs. 25(g) and 25(h)]. In addition, the calculated electronic occupancy of the $zx$ and $yz$ orbitals becomes inequivalent near $E_F$ below $T_N$,[152] thus supporting the above interpretation of polarization-dependent laser-ARPES. Theoretical studies have suggested important roles for such inequivalent electronic occupation in the $zx$ and $yz$ orbitals, i.e., orbital ordering, as the third degrees of freedom characterizing the ground state of the parent compounds of the iron-based superconductors.[153–155]

*4.2.2 Superconducting gap symmetry in optimally doped $BaFe_2(As,P)_2$*

In iron-based superconductors, both $T_N$ and $T_s$ decrease in a similar manner upon ion substitution, and a dome-shaped SC phase appears. A high $T_c$ of up to 55 K[156] is not expected from a conventional electron pairing mediated by phonons.[157] The nesting tendency connecting quasi-cylindrical FSs has been considered to be necessary for electron pairing mediated by the AF spin fluctuations.[147,148] On the other hand, it has also been proposed that orbital ordering[153–155] occurs at $T_s$, as also suggested by the laser-ARPES results for Ba122.[152] On the basis of the multiorbital nature of iron-based superconductors, electron pairing through orbital fluctuations has been proposed.[158–161] Experimental studies have favored different SC pairing symmetries. The sign-reversal superconductivity observed by scanning tunneling microscopy of Fe(Te,Se)[162] and in inelastic neutron scattering (INS) measurements of $(Ba,K)Fe_2As_2$[163] is consistent with the $s_\pm$ wave symmetry expected from the spin fluctuation mechanism. The robustness of $T_c$ against impurities in LaFeAs(O,F)[164] suggests $s_{++}$ wave superconductivity, which can be caused by orbital fluctuations. Such strong material dependence in the pairing symmetry can arise from the balance between these two pairing mechanisms because a crossover from $s_\pm$ to $s_{++}$ symmetry may occur depending on the strengths of microscopic electronic parameters.[158–161] However, no experimental evidence has been reported regarding the role of orbital fluctuations in superconductivity in the iron-based compounds.

The electron pairing mechanism will be reflected in the momentum dependence of the SC gap properties. The spin fluctuation mechanism predicts a strongly orbital-dependent SC gap magnitude,[148,165] but orbital fluctuations should dissolve such orbital dependence.[159–161] Because each FS sheet in iron-based superconductors has a distinct $d$-orbital character, probing the FS sheet dependence of the SC gap magnitude will be a crucial test for investigating the contribution of the orbital fluctuations in the electron pairing.

For this purpose, laser-ARPES was employed to investigate optimally doped $BaFe_2(As,P)_2$ (AsP122).[166] Figure 27(a) shows three hole FSs around the BZ center observed by laser-ARPES. Before investigating the SC gap properties, the orbital characters in these hole FSs should be clarified. According to Ref. 167, the outer hole FS in AsP122 is dominantly composed of the $3Z^2-R^2$ orbital ($X$, $Y$, and $Z$ are the coordinates along the crystal axes of the tetragonal setting) near $k_z = \pi$. However, there is an ambiguity in the orbital character in the band calculations at $k_z = \pi$ for the Ba122 family and, as reported in Ref. 168, the $3Z^2-R^2$ orbital disappears and is switched to an $X^2-Y^2$ orbital at $k_z = \pi$. In order to determine the orbital characters of the three hole bands around the BZ center, especially the outer hole band, polarization-dependent laser-ARPES was performed for sample geometries 1 and 2 as indicated in Figs. 27(b) and 27(c), respectively. The momentum cuts corresponding to the 6.994 eV photons were determined to be $k_z \approx \pi$ by using synchrotron-based ARPES with a low photon energy.[166] Each geometry corresponds to the measurement along the momentum cut illustrated in Figs. 27(d) and 27(e). The selection rule for ARPES indicates that orbitals of even parity are detectable only by light polarization of even parity.[3] In both geometries 1 and 2, the *s*-polarization is active for even-parity orbitals with respect to the mirror planes, although the *p*-polarization is the summation of the even and odd parities. By comparing the results of the *s*-polarization for geometries 1 and 2, we can exclude one of the $X^2-Y^2$ and $3Z^2-R^2$ orbital characters according to the table in Fig. 27(f). Momentum distribution curves (MDCs) near $E_F$ obtained from the *s*-polarization [Figs. 27(g) and 27(h)] clearly show the $k_F$ peaks of the outer hole FS in both geometries. We can thus rule out the possibility of the $X^2-Y^2$ orbital character in the outer hole FS of AsP122. This result is consistent with the calculations in Ref. 167, which reports a dominant $3Z^2-R^2$ orbital component in the outer-hole FS near $k_z = \pi$ for AsP122.

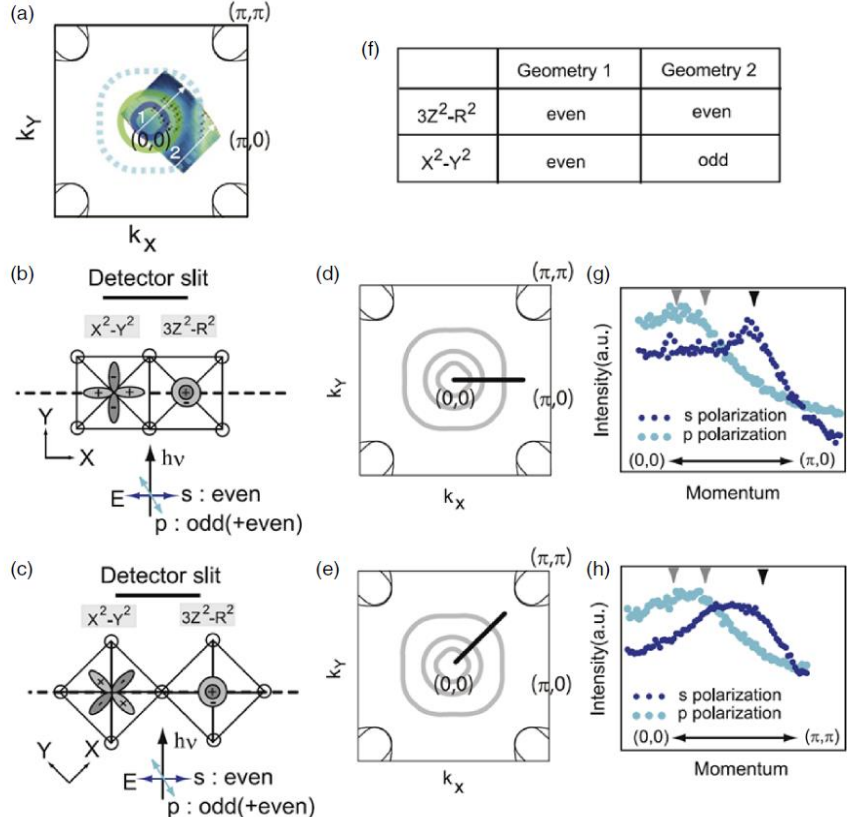

Fig. 27. (Color online) (a) FS mapping around the BZ center ($k_z \sim \pi$) of BaFe$_2$(As$_{0.65}$P$_{0.35}$)$_2$ measured by laser-ARPES. (b), (c) Two experimental configurations for polarization-dependent laser-ARPES. The broken line represents the mirror plane of the crystal. The sample orientation in (b) (geometry 1) is rotated by 45° from that in (c) (geometry 2). (d), (e) Momentum cuts measured in geometries 1 and 2, respectively. (f) Parity of $3Z^2-R^2$ and $X^2-Y^2$ orbitals with respect to the mirror plane in (b) (geometry 1) and (c) (geometry 2). (g), (h) Polarization dependence of the MDCs near $E_F$ along the (0,0)–($\pi$,0) direction (geometry 1) and (0,0)–($\pi$,$\pi$) direction (geometry 2), respectively. Arrows indicate $k_F$ for the inner (gray), middle (gray), and outer (black) hole bands. Reprinted from Ref. 166.

The *E-k* images for optimally doped AsP122 in Figs. 28(a) and 28(b) were obtained by laser-ARPES along cuts 1 and 2 indicated in Fig. 27(a), respectively. The *T* dependence of the EDC at $k_z$ in cut 2 (outer hole FS) shows the SC gap opening below $T_c$ [Fig. 28(c)], and the SC gap magnitude $\Delta$ was directly extracted by a fitting procedure using the BCS spectral function in Eq. (3). The values of $\Delta$ were obtained as a function of $T$ by a fitting procedure that used a BCS function and showed good agreement with the BCS-like $T$ dependence as shown in Fig. 28(d). Figures 28(e)–28(g) show the EDCs in the SC state and the fitting results for each hole FS. As summarized in Fig. 28(h), the FS angle dependence of $\Delta$ indicates the fully gapped nature of the three hole FSs in AsP122. Considering the nodal SC gap symmetry proposed on the basis of magnetic penetration depth[169], thermal conductivity[169,170], and nuclear magnetic resonance measurements[171], our results rule out *d*-wave symmetry and imply the possible presence of line nodes in the electron FSs around the BZ corner. The results of $h\nu$-dependent synchrotron-based ARPES indicated a large anisotropy of the SC gap in electron FSs at the BZ corner, implying possible line nodes.[146] Nearly isotropic and orbital-independent SC gaps have also been observed for the optimally doped sample of BaKFe$_2$As$_2$ (BaK122) by laser-ARPES.[172] The most notable feature commonly observed for AsP122 and BaK122 is the coincidence of $\Delta$ in the three hole FSs [Fig. 29(a)].

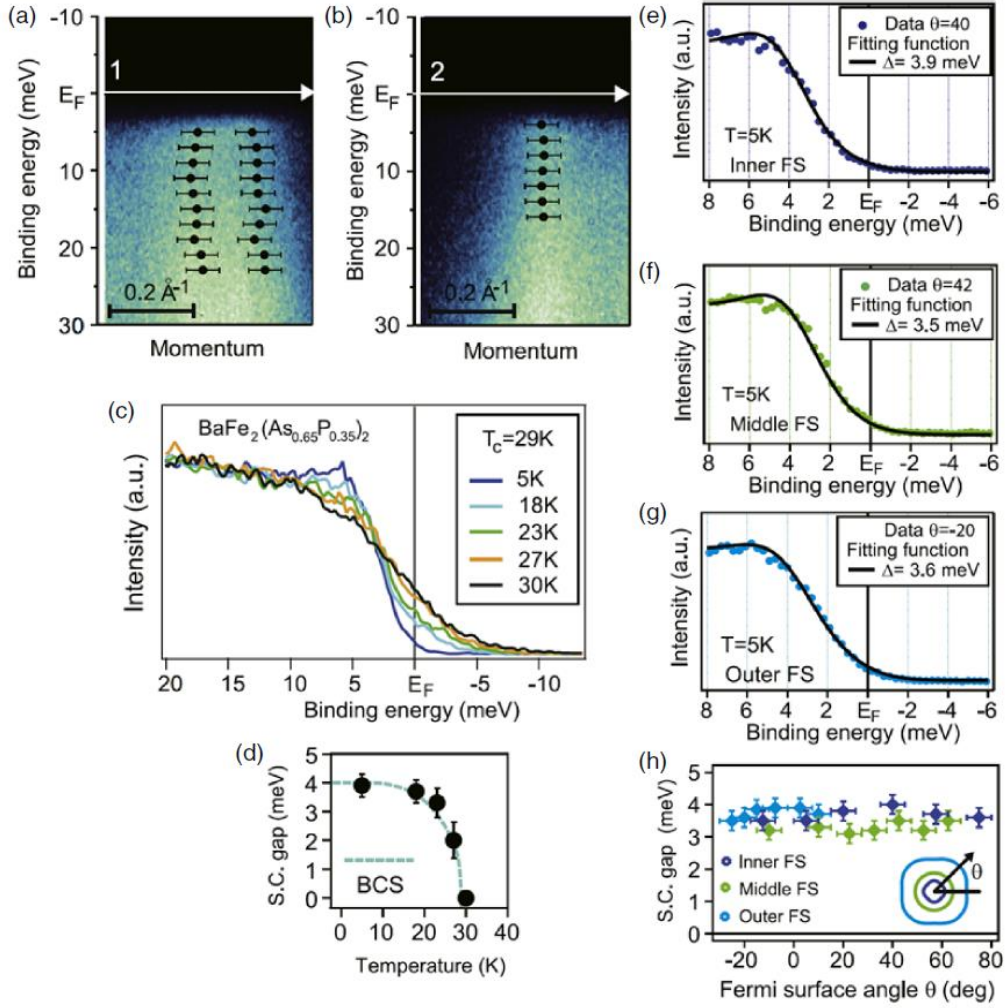

Fig. 28. (Color online) (a), (b) Band dispersions measured by circular polarization along cut 1 and 2 in Fig. 27(a), respectively. (c) $T$ dependence of the EDCs at $k_F$ in cut 2. (d) $T$ dependence of the SC gap magnitude determined by a fitting procedure using a BCS spectral function. The light blue broken curve represents BCS-like $T$ dependence. (e), (g) Fitting results for the EDCs in the SC state measured at $k_F$ for the inner, middle, and outer hole FSs, respectively. (h) FS angle $\theta$ dependence of the SC gap magnitude in each hole FS. The inset shows the definition of $\theta$. Reprinted from Ref. 166.

First-principles band calculations indicate that each hole FS is composed of a different orbital character, as was experimentally confirmed by the strong polarization dependence in three hole bands of BaK122.[172] The comparable $\Delta$ for all hole FSs thus indicates that they are almost orbital-independent.

For the spin-fluctuation pairing mechanism,[147,148,165] the SC gap is closely related to the spin susceptibility. The calculated spin susceptibility suggests dominant electron-pair scattering between disconnected FSs composed of the same orbital character[148,165] (intraorbital pairing), as depicted in Fig. 29(b). Thus, $\Delta$ is expected to be highly sensitive to the orbital character, depending on the overlap of the FS shape. This property, however, contradicts the orbital-insensitive SC gaps observed by laser-ARPES. In particular, the $3Z^2-R^2$ orbital is important because it does not contribute to the electron FSs around the BZ corner in the Ba122 family [Figs. 29(c) and 29(d)].[172] In this case, intraorbital electron pairing between the disconnected FSs is suppressed, and then $\Delta$ for the $3Z^2-R^2$ orbital electrons will be quite small. Taking this into account, the comparable SC gap magnitudes for various orbital characters, including the $3Z^2-R^2$ orbital for the outer hole band, can hardly be explained by the simple spin-fluctuation mechanism.

However, the importance of interorbital coupling[173,174] and the role of the orbital fluctuations[158–161] have been theoretically proposed, which have been neglected in the simple spin fluctuation pairing mechanism. Such interorbital electron pairing is expected to dissolve the orbital dependence of $\Delta$ and can explain the comparable $\Delta$ around the BZ center observed by laser-ARPES. A multiorbital band structure near $E_F$ in iron-based superconductors has the potential to enhance the orbital fluctuations. In addition, the orbital fluctuations are expected to evolve in the vicinity of the orbital ordered state. The laser-ARPES of Ba122 indicated unequal occupation in the electronic structure of $zx$ and $yz$ orbitals below the magnetostructural transition,[152] supporting the orbital ordering scenario.[153-155] While dominantly orbital-fluctuation-derived superconductivity is somewhat speculative and needs to be confirmed by further work, the present laser-ARPES results suggest that the orbital fluctuations should be considered at least on the same footing as the spin fluctuations.

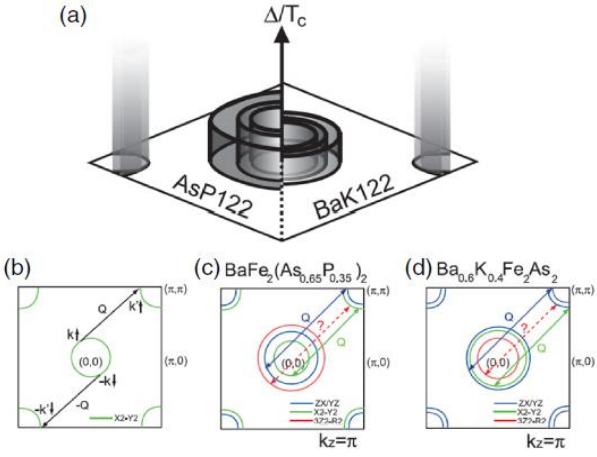

Fig. 29. (Color online) (a) Schematic momentum dependence of the SC gap magnitude around $k_z = \pi$ in AsP122 and BaK122 determined by laser angle-resolved photoemission spectrometry. Those in electron FSs around the BZ corner (light gray) are speculation. (b) Illustration of the electron pair scattering from $(k\uparrow, -k\downarrow)$ to $(k'\uparrow, -k'\downarrow)$ between disconnected FSs of the same orbital character (intraorbital pairing). Electrons are scattered by the AF spin fluctuations, with the wave vector $Q$ derived from the FS nesting. (c), (d) Schematic FS profiles at $k_z = \pi$ for AsP122 and BaK122, respectively. The number of FS sheets and their orbital characters are obtained from first-principles band calculations using experimental atomic positions for $x = 0.4$ ($x = 0.33$) for BaK122 (AsP122) with the Ba122 structure. Arrows with wave vector $Q$ represent the nesting tendency between FS sheets of the same orbital character. Reprinted from Ref. 172.

*4.2.3 Doping-dependent superconducting gaps in BaKFe$_2$As$_2$*

BaK122 is distinguished from other iron-based superconductors because of the persistence of superconductivity in this compound up to the end member K122. The SC gap properties in BaK122 are expected to change with doping from fully opened gaps near the OP region[175–177] to nodal gaps at the end member K122.[178–180] Therefore, BaK122 is an ideal system for studying the doping evolution of SC gap properties. It will also be crucial to investigate the significance of K doping for the low-lying electronic structure in BaK122, which will provide an important background to the theories on iron-based superconductors.

Figures 30(a1)–30(f1) show *E-k* images obtained by laser-ARPES for different doping levels from $x = 0.3$ to $0.7$.[181] Due to symmetry considerations and the polarization direction of the laser, these *E-k* images mainly display the inner hole band around the BZ center.[20,172,182] Whereas a simple BQP dispersion is observed in the overdoped region, similar to the case for cuprates,[10,121] it becomes complicated toward the optimally doped composition. As indicated by the black curves in Fig. 30, the laser-ARPES data show a robust low-energy kink structure (kink 1) in the dispersion that is doping-dependent, with its energy peaking at the optimally doped level ($x \approx 0.4$) and decreasing towards the underdoped and overdoped sides, as indicated by the colored regions in Figs. 30(a2)–30(f2). We attribute kink 1 to the electron-mode

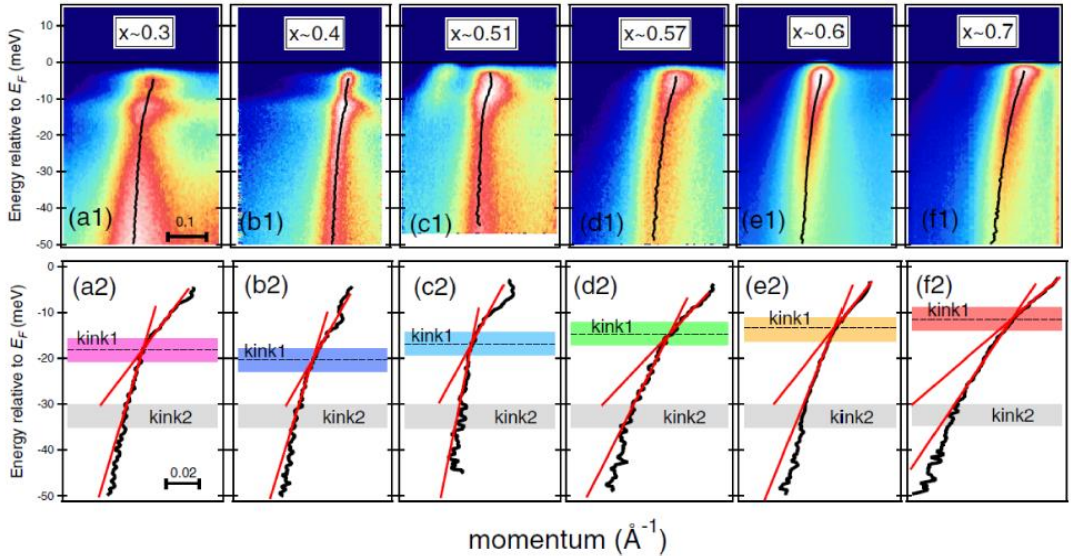

Fig. 30. (Color online) (a1)–(f1) Doping dependence of the *E-k* images for BaK122 obtained by laser-ARPES at a cut along or close to the high-symmetry line in a momentum space measured in the SC state ($T = 7$ K) with a circularly polarized laser. (a2)–(f2) Peak positions of the MDCs (black curves). Red lines are guides to the eye for estimating the kink energies. Kinks 1 and 2 are indicated by colored regions. Reprinted from Ref. 181.

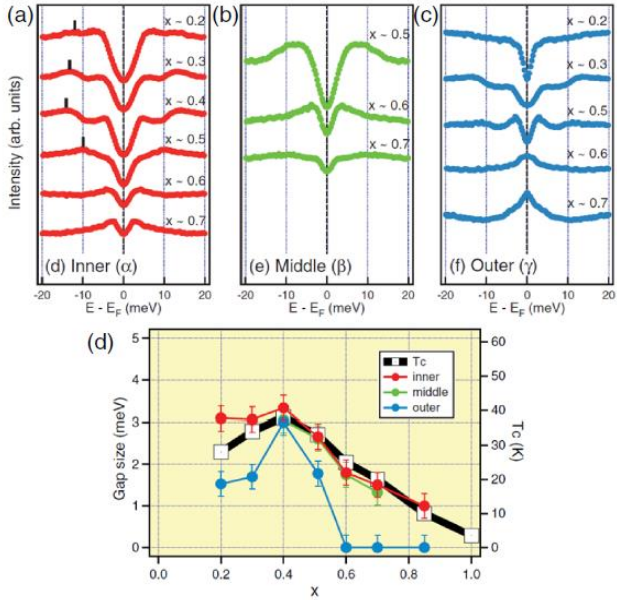

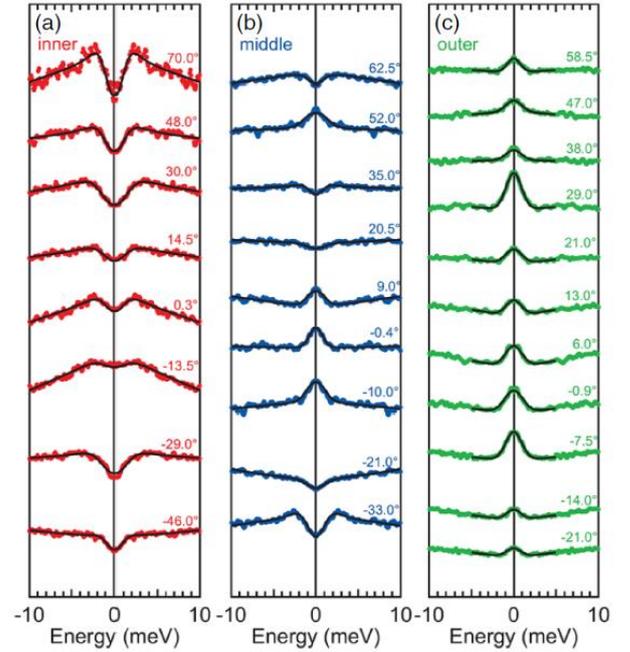

Fig. 31. (Color online) (a)–(c) Doping dependence of the symmetrized EDCs for the inner, middle, and outer hole FSs, respectively. The black bars indicate the position of an additional peak that is distinguished from the SC peak at a lower binding energy. (d) Summary of the doping dependence of $T_c$ (right axis) and the SC gap size (left axis) for the different hole bands of BaK122 around the BZ center. The $x = 0.4$ data were taken from Ref. 172. Reprinted from Ref. 182.

coupling, in good agreement with INS[183] and STS[184] measurements of the same compound, in which a bosonic mode was observed at a similar energy scale, which is considered to be related to spin excitations. The relation between $T_c$ and the mode energy deduced from the laser-ARPES data follows the universal relation.[181] Kink 1 is also $T$-dependent and survives up to ~90 K for the optimally doped case.[181] In addition, we were able to resolve another kink at higher $E_B$ (kink 2) that shows less doping and $T$ dependence than kink 1 and thus may have a different origin.

In order to investigate the doping dependence of the SC gap properties, EDCs obtained at $k_F$ for the inner, middle, and outer hole bands are symmetrized and shown in Figs. 31(a)–31(c), respectively. The symmetrized EDCs suggest that, for all doping levels, the SC gap opens on all the hole bands by showing a dip at $E_F$, except on the outer bands of the heavily overdoped samples ($x = 0.6$ and 0.7), where a peak at $E_F$ is observed [Fig. 31(c)]. The SC gap magnitude was precisely determined by fitting the EDCs to a BCS spectral function; the results are displayed in Fig. 31(d). It becomes clear that by moving away from the optimally doped region, the orbital-independent SC gaps become strongly orbital-dependent. The most striking feature is the abrupt drop in the outer FS sheet ($X^2-Y^2$ orbital) gap size in the heavily overdoped region ($x = 0.6$) while the inner and middle FS ($ZX/YZ$ orbitals) gaps roughly scale with $T_c$. Indeed, this behavior is accompanied by the simultaneous disappearance of the electron FS sheet with similar orbital character ($X^2-Y^2$ orbital) at the BZ corner at $x = 0.6$. These results indicate the different contributions of $X^2-Y^2$ and $ZX/YZ$ orbitals to the superconductivity in BaK122.

Fig. 32. (Color online) (a)–(c) Symmetrized EDCs at $k_F$ of inner, middle, and outer FSs, respectively. The data were obtained at 2.0 K (below $T_c$) for various FS angles $\varphi$ as shown in each panel. Solid curves represent the fitting functions using a Dynes function. Reprinted from Ref. 20.

### 4.2.4 Superconducting gap nodes in $KFe_2As_2$

K122 is the end member of hole-doped BaK122, and there is experimental evidence for the existence of SC gap nodes.[178,179,185] Extreme hole doping changes the electron sheets into cloverlike hole pockets around the BZ corner, which makes this material different from other iron-based systems showing possible nodal superconductivity, such as LaFePO[186] and AsP122.[146,170,187] The characteristic FS topology in K122 has led to a recent theoretical suggestion of $d$-wave superconductivity with gap nodes in all the zone-centered hole bands,[188,189] which would imply a phase transition with a change in the SC gap symmetry from $s$- to $d$-wave upon changing the doping level. In support of this scenario, thermal conductivity measurements have suggested possible $d$-wave gap symmetry in K122.[190] It is therefore essential to determine the SC gap symmetry and the positions of the nodes in K122.

Shown in Figs. 32(a)–32(c) are the symmetrized EDCs at $k_F$ for the inner, middle, and outer FSs, respectively, obtained at 2.0 K (below $T_c$) by employing the second-generation laser-ARPES.[20] Each symmetrized EDC at $k_F$ is identified with an FS angle $\varphi$. In contrast to the case for the inner FS, which shows a clear gap in the symmetrized EDCs for all $\varphi$, the spectra in the middle FS exhibit a peak or a valley at $E_F$ as a function of $\varphi$, indicative of highly anisotropic SC gap symmetry with nodes. On the other hand, the symmetrized EDCs in the outer FS only show a peak at $E_F$ for all $\varphi$, indicative of a negligible gap in the entire outer hole FS. FS-sheet-dependent SC gaps were thus clearly observed for K122, which is different from the orbital-independent SC gaps for optimally doped BaK122,[172] but is smoothly connected to the evolution of the SC gaps as a function of K-doping.[182] The magnitude of the SC gaps was quantitatively estimated as a function of $\varphi$ through a fitting procedure using a Dynes function,[24] as summarized in Fig. 33(a). Because the SC coherent peak is not sharp for all $k_F$ and the number of fitting parameters for the BCS spectral function is more than that for the Dynes function, the fitting parameters did not converge to reasonable values. We thus used the Dynes function to obtain the SC gap magnitude of K122. The solid circles in Fig. 33(a) are derived from the symmetrized EDCs in Fig. 32. Open circles are also plotted by assuming the tetragonal crystal symmetry. In order to resolve the nodal position near $\varphi = 0°$, the SC gaps in the middle FS were further investigated in detail. As shown in the inset of Fig. 33(a), the nodal positions were found to be around $\varphi \approx \pm 5°$.

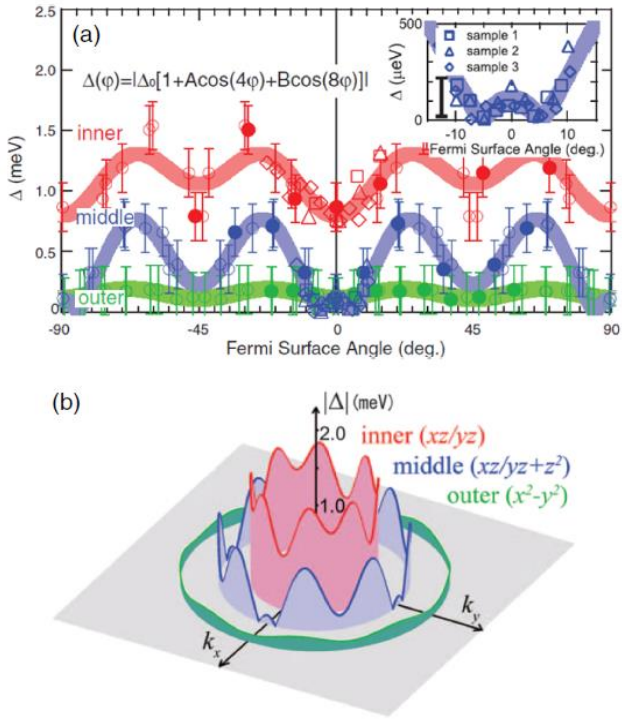

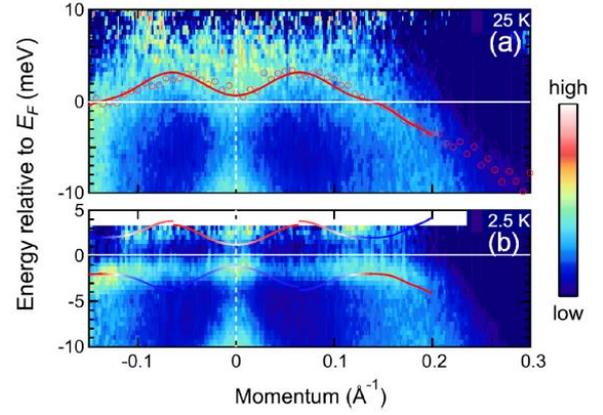

Fig. 34. (Color online) (a), (b) E-k images for FeTe$_{0.6}$Se$_{0.4}$ taken above and below $T_c$, respectively. Reprinted from Ref. 211.

Fig. 33. (Color online) (a) Summary of the SC gap size for three hole FSs plotted as a function of FS angle $\varphi$. The solid circles are determined by the fitting procedure. The open circles are plotted by symmetrizing, taking into account the tetragonal crystal symmetry. The SC gap sizes of the inner and middle FSs obtained from additional high-resolution measurements around $\varphi \sim 0°$ are also plotted using open rectangles, triangles, and diamonds. The curves on the markers are the model SC gap functions for the inner, middle, and outer hole FSs, respectively. The width of these lines corresponds to the standard deviations of the fitting parameter $\Delta$ of each FS sheet. The inset shows SC gap sizes around $\varphi = 0°$ for the middle FS with an enlarged scale. (b) Three-dimensional plot of the SC gap sizes. Reprinted from Ref. 20.

We can discuss some implications of the variety in the SC gap anisotropy for three hole FSs. First of all, the full-gap nature of the inner FS is consistent with the s-wave symmetry and excludes the d-wave symmetry as suggested by theoretical[188,189] and experimental studies.[190] Second, the SC gap symmetries of the inner and middle FSs show strong anisotropy, whereas that on the outer FS is negligibly small. The SC gap anisotropy was further analyzed using the two lowest harmonics under the fourfold tetragonal symmetry as follows:

$$\Delta(\varphi) = [1 + A\cos(4\varphi) + B\cos(8\varphi)]. \quad (7)$$

Such a $\cos(4\varphi)$ term has been considered in some theoretical studies.[191,192] In the present result, the higher-order $\cos(8\varphi)$ term is required to reproduce the gap minima observed around $0°$ and $45°$. This simple analysis successfully explains the experimental anisotropy in the SC gaps, as fitted by the thick curves in Fig. 33(a). Figure 33(b) depicts the obtained SC gap symmetry for each hole FS sheet. The nodal points in the middle FS are determined to be $\varphi = \pm(5.34 \pm 0.04)°$. Finally, these gap functions are consistent with the London penetration depth results.[20] Thus, the present laser-ARPES results suggest that K122 is a nodal s-wave multigap superconductor with unusual eightfold sign reversal in its gap function [Fig. 33(b)]. This implies that in spite of the large variety of gap functions even within the Ba122 family, the basic symmetry of the SC order parameter always belongs to the $A_{1g}$ representation, which is symmetrical with respect to fourfold rotation.[193] Systematic evolution of the nodal position in the heavily K-doped region ($0.76 \leq x \leq 1.0$) has been further resolved by the second-generation laser-ARPES, and multiple electron-pair scattering channels were also suggested from the results.[194]

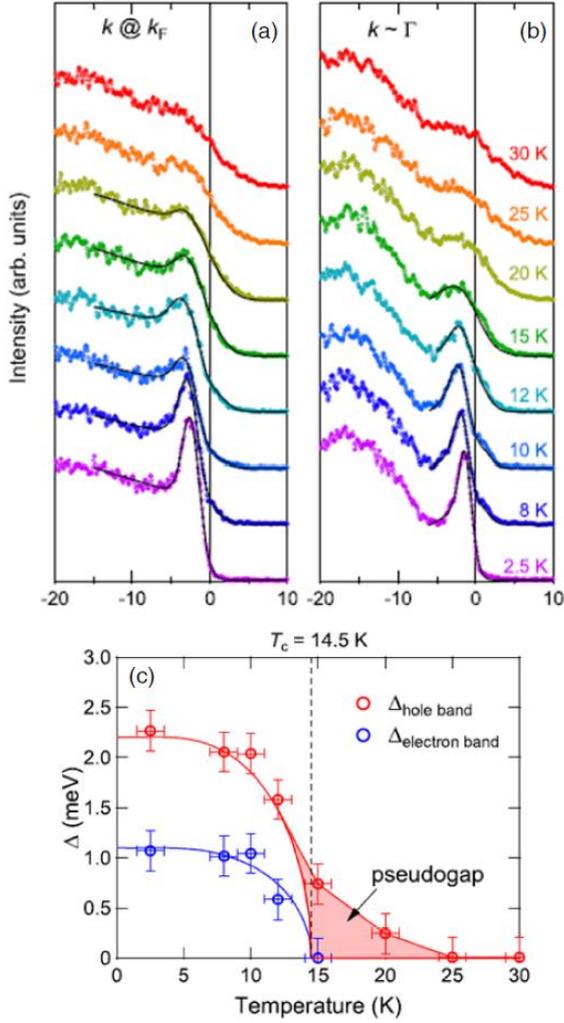

Fig. 35. (Color online) (a), (b) $T$ dependence of the EDCs at $k = k_F$ for the $X^2-Y^2$ hole band and $k = \Gamma$ (bottom of the electron band), respectively. Fitting results with the BCS spectral function are indicated by the black solid curves. (c) $T$ dependence of the obtained SC gap sizes estimated from the fitting procedure. The shaded area represents PG formation for the hole band. Reprinted from Ref. 211.

*4.2.5 Composite BCS–BEC superconductivity in Fe(Te,Se)*

Observation of the PG in the normal state of the high-$T_c$ cuprates leads to a remarkable challenge to our understanding of superconductivity.[195–197] In particular, is the PG related to the preformed Cooper pairs, or does it reflect another ground state competing with superconductivity?[198–200] A PG was also found in a strongly interacting ultracold Fermi gas above the superfluid condensation $T$.[201] These results suggest the importance of the pairing in the normal phase, above the onset of superfluidity/superconductivity in strongly interacting systems. PG formation has been discussed in the scheme of a BCS/BEC crossover.[202] Very interestingly, a study on ultracold Fermi gases reported a PG for a system with $\Delta/\varepsilon_F \approx 1$,[201] while the optimally doped Bi2212 exhibits a value of $\Delta/\varepsilon_F \approx 0.1$,[203] where $\Delta$ and $\varepsilon_F$ represent the SC gap size and the Fermi energy, respectively. A recent ARPES study of an Fe(Te,Se) superconductor reported that the system is in the BCS–BEC crossover regime, with $\Delta/\varepsilon_F \approx 0.5$ associated with a hole band centered at the $\Gamma$ point.[203] While the band dispersions were similar to those in previous ARPES studies of the normal phase,[204,205] $\Delta$ was much smaller than in an earlier report, which claimed strong coupling superconductivity.[206]

In order to investigate the PG in Fe(Te,Se), which is expected from the theories of BCS–BEC crossovers,[207–210] the electronic structure very close to $E_F$ was investigated across $T_c$ by employing the second-generation laser-ARPES.[211] Shown in Figs. 34(a) and 34(b) are the obtained $E-k$ images divided by the FD function above and below $T_c$, respectively. The solid curves in panel (b) are the BQP dispersions depicted using the normal-state dispersions in panel (a) and the $\Delta(k)$ of 2 meV for the hole band and $\Delta(k)$ of 1 meV for the electron band. It is noteworthy that the BQP dispersions merge for the electron and hole bands, suggestive of a composite BCS–BEC superconductivity. This result also suggests that irrespective of the strength of the coupling, both the hole and electron bands in the SC state exhibit BQP dispersions due to particle-hole mixing.[212] However, the superconductivity in the electron band is expected to be very sensitive to the occupancy of the electron band with Se substitution for Te, as well as pressure. This possibly explains the large variation in $T_c$ with pressure for FeSe ($T_c$ = 8.5–36.7 K).[213] Another difference was also found in a comparison between the weak-coupling hole band and the strong-coupling electron band. Figures 35(a) and 35(b) show the $T$ dependence of the EDCs at $k_F$ for the hole band and $k = \Gamma$ for the electron band, respectively. The black solid curves indicate the fitting results using a BCS spectral function. The SC gap magnitude is estimated and summarized in Fig. 35(c). Intriguingly, PG formation was detected only for the weaker-coupling hole band in the spectra above $T_c$. However, in strong contrast to the BCS–BEC crossover theory,[210] which predicts the existence of a PG in the BEC strong-coupling regime, the electron band does not show a PG above $T_c$. The coexistence of weak- and strong-coupling superconductivity was observed in the same material, which is not fully consistent with the present understanding of weak- and strong-coupling superconductivity.

## 5. Summary

We reviewed the developments in the laser-PES system at Institute for Solid State Physics at the University of Tokyo. A combination of high energy resolution, high cooling ability, and bulk sensitivity has enabled a wide range of applications, highlighting the laser-PES as a versatile probe to investigate fine electronic structures. Importantly, investigations of unconventional superconductivity are possible for extremely low-$T_c$ superconductors ($T_c > 1.5$ K). At present, high-energy-resolution laser-PES is a widespread method, and a growing number of studies involving its use have been reported.[214-217] On the other hand, the laser technique for PES has been extended to the pump-probe method using a femtosecond-pulse laser. Recent developments in higher-order harmonic generation with a pulsed gas jet enable the use of coherent photons for PES with $h\nu$ up to 60 eV.[218] Advances in the laser technique will be essential for developing a sophisticated PES method and achieving epoch-making breakthroughs in the field of solid state physics.

**Acknowledgments**

T. Kiss, K. Ishizaka, Y. Ota, Y. Kotani, K. Kaji, T. Kondo, W. Malaeb, M. Okawa, F. Kanetaka, T. Togashi, T. Kanai, X.-Y. Wang, C.-T. Chen, and S. Watanabe are gratefully acknowledged for developing the laser-PES system at Institute for Solid State Physics at the University of Tokyo. We thank S. Tsuda, R. Eguchi, Y. Ishida, A. Chainani, T. Yokoya, A. Shimoyamada, and Y. Shibata for valuable discussions.